\documentclass[11pt]{article}
\usepackage{amsmath, 
            amssymb, 
            amsthm,bm,
            bbm}
\usepackage{geometry}
\RequirePackage[colorlinks,citecolor=blue]{hyperref}

\usepackage{setspace}

\usepackage{algorithm}
\usepackage{algpseudocode}

\RequirePackage[numbers]{natbib}

\usepackage{multirow}
\usepackage{booktabs}
\usepackage{orcidlink}
\usepackage{pdflscape}

\theoremstyle{definition}

\usepackage{textalpha}

\title{\bf Mixture-of-experts Wishart model for covariance matrices 
with an application to Cancer drug screening
}
\author{The Tien Mai$^{1}$\orcidlink{0000-0002-3514-9636}
,
Zhi Zhao$^{2}$\orcidlink{0000-0003-2325-1438}
}

\date{
\small
$^{1}$Norwegian Institute of Public Health, 0456 Oslo, Norway.
\\
$^{2}$University of Oslo, 0372 Oslo, Norway.
\\
email: the.tien.mai@fhi.no
}

\begin{document}

\maketitle

\begin{abstract}
Covariance matrices arise naturally in different scientific fields, including finance, genomics, and neuroscience,
where they encode dependence structures and reveal essential features of complex multivariate systems.
In this work, we introduce a comprehensive Bayesian framework for analyzing heterogeneous covariance data through both classical mixture models and a novel mixture-of-experts Wishart (MoE–Wishart) model. The proposed MoE–Wishart model extends standard Wishart mixtures by allowing mixture weights to depend on predictors through a multinomial logistic gating network. This formulation enables the model to capture complex, nonlinear heterogeneity in covariance structures and to adapt subpopulation membership probabilities to covariate-dependent patterns. To perform inference, we develop an efficient Gibbs-within–Metropolis–Hastings sampling algorithm tailored to the geometry of the Wishart likelihood and the gating network. We additionally derive an Expectation–Maximization algorithm for maximum likelihood estimation in the mixture-of-experts setting.
Extensive simulation studies demonstrate that the proposed Bayesian and maximum likelihood estimators achieve accurate subpopulation recovery and estimation under a range of heterogeneous covariance scenarios. Finally, we present an innovative application of our methodology to a challenging dataset: cancer drug sensitivity profiles, illustrating the ability of the MoE–Wishart model to leverage  covariance across drug dosages and replicate measurements.

Our methods are implemented in the \texttt{R} package \texttt{moewishart} available at  \url{https://github.com/zhizuio/moewishart}. 

\end{abstract}


\section{Introduction}

In traditional statistical modeling, it is often assumed that a single, universal relationship between variables holds across an entire population. Observations that deviate from this relationship are typically treated as outliers. In practice, however, real-world systems are rarely homogeneous. Instead, data frequently arise from multiple latent subpopulations, each characterized by distinct behavioral patterns or structural relationships. 
As the scale and complexity of modern datasets continue to increase, it becomes essential to appropriately model such hidden heterogeneity. Mixture models \citep{quandt1978estimating} provide a powerful and principled framework for extending classical clustering and regression methods to settings in which population structure is intrinsically heterogeneous \citep{fruhwirth2019handbook}.

In general,
covariance matrices play a central role in scientific inquiry, as they encode patterns of linear dependence among variables and reveal the underlying structure of multivariate phenomena. Their importance spans a wide range of application domains. In finance, covariance matrices underpin portfolio allocation, risk assessment, and investment strategies \citep{sun2019improved,johansson2023simple}. 
In imaging/pattern recognition, covariance among pixel/voxel intensities or regions captures spatial and spatio-temporal dependencies, supporting tasks as clustering and feature extraction \citep{Tuzel2006}.
In genomics, quantifying gene–gene interactions may help identify biologically meaningful trait associations \citep{schafer2005shrinkage}. In neuroscience, covariance matrices are used to characterize functional connectivity and network-level interactions between brain regions \citep{nielsen2017modeling}. Moreover, covariance structures form the computational foundation of essential multivariate techniques such as Principal Component Analysis and Factor Analysis \citep{runcie2013dissecting,gao2014multiple}.

Despite their ubiquity, covariance structures are often overlooked in standard clustering techniques. Widely used algorithms such as 
k-means and hierarchical clustering primarily rely on differences in mean structure, thereby ignoring potentially rich information encoded in covariance patterns. To address this limitation, recent research has increasingly focused on clustering based on covariance heterogeneity. An important probabilistic framework was introduced by \cite{Hidot2010}, who proposed a mixture model with Wishart component densities and an Expectation–Maximization (EM) algorithm for parameter estimation. Subsequent extensions include sparse penalization for high-dimensional covariance estimation \citep{cappozzo2025model}, variational Bayesian inference \citep{nielsen2017modeling}, and applications to discriminant analysis \citep{chaudhari2021discrimination}. Relatedly, \cite{Gallaugher2018} considered finite mixtures of skewed matrix-variate distributions, including the matrix-variate variance–gamma distribution as a generalization of the Wishart, with inference carried out via EM algorithms. Despite these advances, a comprehensive fully Bayesian treatment of Wishart mixture models has received relatively limited attention.

In this paper, we address this gap by first developing a fully Bayesian formulation of the classical mixture of Wishart distributions. Building on this foundation, we introduce a novel Mixture-of-Experts Wishart (MoE–Wishart) model, which integrates the flexibility of mixture-of-experts architectures with the interpretability and structure of Wishart-distributed covariance data. Unlike standard mixture models with fixed mixing proportions, mixture-of-experts models \citep{jordan1994hierarchical} allow cluster membership probabilities to depend explicitly on covariates, thereby enabling the model to learn how latent subpopulations vary across the predictor space. Through a probabilistic gating network governed by a softmax link \citep{peng1996bayesian}, the MoE framework partitions the input space and dynamically assigns observations to expert components. This structure allows the proposed model to capture complex, nonlinear, and piecewise behavior in covariance patterns, which is particularly valuable in high-dimensional or highly heterogeneous settings. For a general overview of mixture-of-experts models, see \cite{nguyen2018practical}.

For posterior inference in both the  mixture model and the proposed MoE–Wishart extension, we utilize efficient Markov chain Monte Carlo (MCMC) algorithms. In particular, we construct a Gibbs-within–Metropolis–Hastings sampler tailored to the geometry of the posterior distribution induced by symmetric positive definite covariance matrices. In addition, we derive an Expectation–Maximization algorithm for maximum likelihood estimation in the mixture-of-experts setting, where closed-form updates are not readily available.

The main contributions of this paper are therefore threefold: (i) a fully Bayesian formulation of the mixture of Wishart distributions, (ii) the introduction of a new mixture-of-experts model for covariance matrix data (MoE–Wishart model), and (iii) tailored inference algorithms for the MoE–Wishart model, including both MCMC and EM procedures. 
To facilitate reproducibility and practical adoption, we additionally provide an R package \texttt{moewishart} that implements all proposed methods (available at  \url{https://github.com/zhizuio/moewishart}).

We conduct extensive simulation studies to assess the statistical accuracy and computational performance of the proposed methods. Finally, we illustrate their practical utility through a real-data application to study covariance patterns arising in cancer drug discovery and repurposing.
In biomedicine, drug repositioning refers to finding new uses for approved and preclinical drugs \citep{Ashburn2004}. 
With the growing availability of data from large-scale drug screens and relevant molecular information, data-driven drug repositioning has the potential to accelerate clinical advances. 
Specifically, exploiting drug–drug similarity or clusters based on shared molecular targets and pharmacological or chemical structures can facilitate drug repurposing \citep{Xue2018,Huang2025}. 
Drug screening data are typically derived from experiments in which individual drugs, tested at multiple dosages/concentrations, are applied to patient tissue samples or cell lines. 
Each drug's efficacy is therefore assessed via cell viability measured on replicated samples across multiple dosages. 
Currently, most models use a scalar summarized descriptor (e.g., IC$_{50}$ or area under the drug dose-response curve) to represent each drug's efficacy \citep{SeashoreLudlow2015,Mai2020,SharifiNoghabi2021,Zhao2022}, but this approach loses substantial information. 
We propose to use, for each drug, the covariance descriptor of its response data (i.e., covariance matrix of cell viability on replicated samples across multiple dosages), and then model latent clusters based on covariance matrices from many drugs. 
Drugs clustered in one group may indicate similar functionality and suggest new uses of some drugs. 

The remainder of the paper is organized as follows. Section~\ref{sc_model_mixture} introduces the Wishart mixture and mixture-of-experts models for covariance matrix data.
Section~\ref{sc_MCMC_algorithms} develops a Bayesian framework together with MCMC sampling algorithms for both the mixture and mixture-of-experts models.
Section~\ref{sc_MLE_EM} presents an EM algorithm for obtaining maximum likelihood estimates in the mixture-of-experts model. 
 Simulation studies are reported in Section~\ref{sc_simulations} and
an application to cancer drug screening data is presented in Section~\ref{sc_application_drug}.
We discuss and conclude in Section \ref{sc_conclusion}.

\section{Mixture of covariance matrices models}
\label{sc_model_mixture}

We first describe the mixture model and then present extension to mixture-of-expert models.

\subsection{Mixture model}

Let \(S_1,\dots,S_n\) be observed symmetric positive definite (SPD) \(p\times p\) matrices.  We model these as a finite mixture of Wishart distributions:
\begin{equation}
    \label{eq_mixture_model}
    S_i \mid z_i=k, \Sigma_k, \nu_k \;\overset{\text{ind}}{\sim}\; \mathcal{W}_p(\nu_k, \Sigma_k),
\qquad i=1,\dots,n,
\end{equation}
where \(z_i\in\{1,\dots,K\}\) denotes the latent (unobserved) cluster label for observation \(i\).
Here, \(\nu_k>p-1\) is the degrees of freedom of the Wishart, and \(\Sigma_k\) is the cluster-specific scale (covariance) matrix for cluster \(k\). 
The Wishart density (with our parametrization) is
\begin{equation}
    f_{\mathcal W}(S \mid \nu,\Sigma)
= 
\frac{|S|^{(\nu-p-1)/2} \exp\left(-\frac{1}{2}\operatorname{tr}\!\left(\Sigma^{-1} S\right)\right)}
{2^{\nu p/2} |\Sigma|^{\nu/2} \,\Gamma_p\!\left(\frac{\nu}{2}\right)}
,
\label{eq_density_of_wishart}
\end{equation}
where \(\Gamma_p(\cdot)\) is the multivariate Gamma function.

Marginalizing over the latent cluster labels \(z_i\), the distribution of each observed matrix \(S_i\) can be written in a mixture density form. 
Let \(\boldsymbol{\pi}=(\pi_1,\dots,\pi_K)\) denote the (fixed) mixture weights, where 
\(\pi_k = \Pr(z_i = k)\), \(\pi_k \ge 0\), and 
\(\sum_{k=1}^K \pi_k = 1\). Then the conditional density of covariance observation \(S_i\) (marginal over the latent cluster labels $z_i$) is 
 \begin{equation}
 \label{eq_mixture_model_02}
 f(S_i \mid \boldsymbol{\pi}, 
 \{\Sigma_k,\nu_k\}_{k=1}^K)
 =
 \sum_{k=1}^K \pi_k 
 f_{\mathcal W} \left(S_i \mid \nu_k, \Sigma_k\right),
 \qquad i=1,\dots,n,
 \end{equation}
where \(f_{\mathcal W}(\cdot \mid \nu_k,\Sigma_k)\) is the Wishart density defined in \eqref{eq_density_of_wishart}. 
This formulation highlights that the population distribution of SPD matrices is represented as a convex combination of \(K\) Wishart components, each corresponding to a distinct latent cluster with its own scale matrix and degrees of freedom. 
Such finite mixture of Wishart distributions offers a coherent and interpretable framework for modeling heterogeneous collections of covariance matrices, while explicitly respecting their geometric and positive-definiteness constraints. By allowing both the scale matrices and degrees of freedom to vary across clusters, the model captures differences in both covariance structure and dispersion, yielding flexible yet parsimonious representations of latent subpopulations.

\subsection{Mixture–of–Experts Wishart model}

Let $S_1,\dots,S_n$ be observed SPD matrices of size $p\times p$ and 
let $X\in\mathbb{R}^{n\times q}$ be additional covariates.  
We consider an extension of the mixture model in \eqref{eq_mixture_model} to a finite mixture-of-experts (MoE) model with $K$ Wishart experts (MoE–Wishart):
\begin{equation*}
 \begin{aligned}
    S_i \mid z_i=k, \Sigma_k, \nu_k 
   & \;\overset{\text{ind}}{\sim}\; \mathcal{W}_p(\nu_k,\Sigma_k),
\\
\Pr(z_i=k\mid X_i) 
& = \pi_{ik},
\quad
\text{ with }
k = 1, \ldots, K,
\end{aligned}      
\end{equation*}
where the gating probabilities are multinomial logistic (softmax) functions of $X_i$:
\begin{equation}
    \pi_{ik} (X_i;\beta_k)
=
\frac{\exp(X_i^\top\beta_k)}{\sum_{\ell=1}^K \exp(X_i^\top\beta_\ell)} ,
\qquad \beta_K\equiv 0\ 
\text{(for identifiability)}.
\end{equation}
Under this MoE formulation, the conditional density of $S_i$ given covariates $X_i$ (marginal over the latent cluster labels $z_i$) is
\begin{equation}
       \label{eq_MoE_model}
f(S_i \mid X_i,  \{\beta_k,\Sigma_k,\nu_k\}_{k=1}^K)
=
\sum_{k=1}^K \pi_{ik}(X_i;\beta) \,
f_{\mathcal W}\!\left(S_i \mid \nu_k, \Sigma_k\right).
\end{equation}

This MoE–Wishart model extends the baseline finite mixture formulation by explicitly incorporating covariate information into the clustering mechanism, thereby allowing cluster membership to vary systematically with observed predictors rather than being governed by fixed mixing proportions. 
In the literature, $X_i$ are also called concomitant variables, so that the MoE models are also referred to as mixture models with concomitant covariates. 
While the previous mixture model assumes exchangeable observations with homogeneous prior cluster probabilities, the MoE formulation introduces a multinomial logistic gating network that adapts the allocation probabilities to subject-specific covariates, yielding a more flexible and interpretable representation of heterogeneity. In this setting, the Wishart experts retain their role in modeling cluster-specific covariance structure and dispersion through 
\((\Sigma_k,\nu_k)\), but the inclusion of \(X_i\) enables the model to capture covariate-driven shifts in latent regimes, which is particularly important in applications where structural variability is partially explained by external factors.
When $X_i$ is the $\bm 1$-column, {i.e.}, the intercept, the MoE model reduces to the mixture model.

The identifiability of the MoE–Wishart model follows from general results for  mixtures of exponential family experts \citep{jiang1999identifiability}. Specifically, the model is identifiable up to a permutation of the latent cluster labels, provided the cluster-specific Wishart distributions are distinct and the covariate design matrix is of full rank. The constraint $ \beta_K\equiv 0 $
 resolves the translational non-identifiability inherent to the multinomial logistic gating network, ensuring unique parameter estimation for a given labeling of the mixture components.

\section{Bayesian approaches and MCMC algorithms}
\label{sc_MCMC_algorithms}

In this section, we introduce a Bayesian framework together with MCMC sampling algorithms specifically designed for the mixture and MoE–Wishart models. 
To the best of our knowledge, this constitutes a novel contribution.

\subsection{Bayesian mixture–Wishart model}

We place the following priors distribution on the parameters in the mixture model in \eqref{eq_mixture_model_02}:
\begin{align*}
z_i \mid \bm\pi &\sim \operatorname{Categorical}(\pi_1,\dots,\pi_K), \\
\bm\pi &\sim \operatorname{Dirichlet}(\alpha_1,\dots,\alpha_K),\\
\Sigma_k &\sim \mathcal{IW}(\nu_0, \Psi_0^{-1}), \qquad k=1,\dots,K,
\end{align*}
where \(\mathcal{IW}(\nu_0,\Psi_0^{-1})\) denotes the inverse–Wishart prior on \(\Sigma_k\) with degrees of freedom \(\nu_0\) and scale matrix \(\Psi_0^{-1}\) (we follow the common conjugate parametrization where the posterior for \(\Sigma_k\) is inverse–Wishart with updated parameters).  For the cluster Wishart degrees \(\nu_k\) we place an independent Gamma prior:
\[
p(\nu_k) \propto \nu_k^{a-1} e^{-b \nu_k},\qquad \nu_k>0,
\]
with hyper-parameters \(a>0, b>0\).  In practice the sampler enforces \(\nu_k>p-1\).
The Dirichlet hyperparameters 
\((\alpha_1,\dots,\alpha_K)\) control prior concentration of the mixture weights,  we use a symmetric specification 
\(\alpha_k=\alpha_0/K\) by default, where smaller 
\(\alpha_0\) favors sparse component usage and larger
\(\alpha_0\) favors near-uniform weights. For the inverse–Wishart prior, 
\(\nu_0\) determines prior strength and is set to 
\(\nu_0=p+2\) for a weakly informative prior with finite mean, while 
\(\Psi_0\) is chosen proportional to the identity matrix. The Gamma hyperparameters 
\((a,b)\) for \(\nu_k\) are selected so the prior mean 
\(a/b\) lies moderately above \(p-1\), ensuring validity while remaining weakly informative.

Let \(\Theta = \{\bm\pi, z_{1:n}, (\nu_k,\Sigma_k)_{k=1}^K \} \).
The joint posterior distribution is
\[
p(\Theta \mid S_{1:n})
\propto
\Bigg[
\prod_{i=1}^n \prod_{k=1}^K
\big\{
\pi_k f_{\mathcal W}(S_i\mid \nu_k,\Sigma_k)
\big\}^{\mathbf{1}_{\{z_i=k\}}}
\Bigg]
p(\bm\pi)
\prod_{k=1}^K p(\Sigma_k)p(\nu_k).
\]
Conditioning on the latent labels yields a factorized posterior structure that enables efficient Gibbs sampling with a Metropolis–Hastings step for \(\nu_k\).

\noindent Below we give each update used in the Gibbs sampler and the Metropolis--Hastings (MH) step for \(\nu_k\).
The entire procedure is summarized in Algorithm \ref{alg1}.

{\bf Step 1: Conditional of the labels \(z_i\).}

The posterior probability that \(z_i=k\) (up to normalization) is
\begin{equation}
    \label{eq_conditional_of_z_i}
    p(z_i=k\mid S_i,\{\Sigma_\ell,\nu_\ell\},\bm\pi) \propto
\pi_k \; f_{\mathcal W}(S_i\mid \nu_k,\Sigma_k).
\end{equation}
Hence compute the log-unnormalized weights
\begin{equation}
\label{eq_ell_ik}
   \ell_{ik} = \log\pi_k + \frac{\nu_k-p-1}{2}\log|S_i|
- \frac{1}{2}\operatorname{tr}\!\left(\Sigma_k^{-1} S_i\right)
- \frac{\nu_k p}{2}\log 2
- \frac{\nu_k}{2}\log|\Sigma_k|
- \log\Gamma_p\!\left(\frac{\nu_k}{2}\right), 
\end{equation}
and sample \(z_i\) from categorical probabilities proportional to \(\ell_{ik}\), using \(\exp(\ell_{ik}-\max_k \ell_{ik})\) instead for numerical stability.  
These $\ell_{ik}$ are used both to sample labels and to compute the observed-data log-likelihood for monitoring.
Note that by integrating out $ \bm\pi $ we have 
$
p(z_i=k\mid S_i,\{\Sigma_\ell,\nu_\ell\}, \bm\pi) 
\propto 
f_{\mathcal W}(S_i\mid \nu_k,\Sigma_k) \cdot (\alpha_k+n_{-i,k}) 
$,
see Appendix \ref{secAppendix1} for details, which may make Gibbs sampling mixing better.

{\bf Step 2: Conditional of the weights \(\bm\pi\).}

Given counts \(n_k\) the posterior for \(\bm\pi\) is Dirichlet distributed:
\[
\bm\pi \mid \mathbf z \sim \operatorname{Dirichlet}(\alpha_1 + n_1,\dots,\alpha_K + n_K).
\]

{\bf Step 3: Conditional of the cluster scale matrices \(\Sigma_k\).}

Given assignments \(\{z_i\}\) and the cluster degrees \(\nu_k\), the prior \(\Sigma_k \sim \mathcal{IW}(\nu_0,\Psi_0^{-1})\) is conjugate with the Wishart likelihood.  The posterior is an inverse–Wishart:
\[
\Sigma_k \mid \{S_i: z_i=k\}, \nu_k
\sim 
\mathcal{IW}\Big(\nu_0 + n_k \nu_k,\; \Psi_0^{-1} + S_{(k)} \Big),
\] 
i.e. degrees \(\nu_0 + n_k \nu_k\) and scale matrix \(\Psi_0^{-1} + S_{(k)}\), 
where $S_{(k)}=\sum_{i\in I_k}S_i$ and $I_k=\{i:z_i=k\}$. 
Note: when \(n_k=0\) (i.e., the cluster $k$ diminishes) we sample \(\Sigma_k\) from the prior \(\mathcal{IW}(\nu_0, \Psi_0^{-1})\).

{\bf Step 4: Update of \(\nu_k\) (MH on log scale).}
\\
The degrees of freedom \(\nu_k\) do not admit a convenient conjugate update; we update each \(\nu_k\) using a MH step proposing on the log scale.  Let the current value be \(\nu\) and propose
\[
\log\nu^\ast \sim \mathcal{N}(\log\nu, \sigma_{\text{MH}}^2),
\qquad \nu^\ast = \exp(\log\nu^\ast).
\]
The log posterior for \(\nu\) (for cluster \(k\))—using the cluster-specific summarized quantities \(n_k\), \(S_{(k)}\), and \(L_{(k)}=\sum_{i\in I_k}\log|S_i|\)—is, up to an additive constant,
\begin{align*}
&\log p(\nu \mid \{S_i\}_{i\in I_k}, \Sigma_k)
\\
&= \sum_{i\in I_k} \log f_{\mathcal W}(S_i\mid \nu,\Sigma_k) + \log p(\nu)
\\
&= \frac{\nu-p-1}{2} L_{(k)} - \frac{1}{2}\operatorname{tr}\!\big(\Sigma_k^{-1} S_{(k)}\big)
- n_k\Big(\frac{\nu p}{2}\log 2 + \frac{\nu}{2}\log|\Sigma_k| + \log\Gamma_p\!\big(\tfrac{\nu}{2}\big)\Big)\\
&\quad + (a-1)\log\nu - \nu b
.
\end{align*}
If we denote by \(\ell(\nu)\) the log-posterior, then the acceptance probability for the proposal \(\nu^\ast\) is
\[
\alpha = \min\left\{1,\ \exp\big( \ell(\nu^\ast) + \log\nu^\ast - \ell(\nu) - \log\nu \big) \right\},
\]
where the extra \(\log\nu\) terms are the Jacobian corrections.  
 We only attempt the MH update for proposals with \(\nu^\ast > p-1\).

\begin{algorithm}[!ht]
\caption{Gibbs sampler for Wishart mixture with \(\nu_k\) via MH
\label{alg1}}
\begin{algorithmic}[1]
\State \textbf{Inputs:} \(S_1,\dots,S_n\), number of clusters \(K\), hyperparameters \(\alpha, \nu_0, \Psi_0\), MH proposal scale \(\sigma_{\text{MH}}\), iterations \(N\), burn-in, thin.
\State \textbf{Initialize:} Vectorize \(S_i\) to form \(S_{\mathrm{mat}}\) (size \(n\times p^2\)); compute \(\log|S_i|\) for all \(i\). Initialize labels \(z_i\) (k-means on \(\operatorname{vec}(S_i)\) suggested), \(\pi_k = n_k/n\), \(\nu_k\) (e.g.,\ \(p+2\)), and \(\Sigma_k\) (cluster sample / prior).
\For{iter $= 1,\dots,N$}
  \State \(\triangleright\) \textbf{Step 1 (labels):}
  For each cluster \(k\) compute \(\Sigma_k^{-1}\), \(\log|\Sigma_k|\)
  and \(\operatorname{tr}(\Sigma_k^{-1} S_i)\), and draw labels from
$$
p(z_i=k\mid S_i,\{\Sigma_\ell,\nu_\ell\}, \bm\pi) 
\propto 
f_{\mathcal W}(S_i\mid \nu_k,\Sigma_k) \cdot (\alpha_k+n_{-i,k}) 
$$

\State \(\triangleright\) \textbf{Step 2 (weights):}
Set \(n_k = \#\{ i \,:\, z_i = k \}\) and draw
$$
\bm{\pi} 
\sim
\operatorname{Dirichlet}(\alpha_1 + n_1, \dots, \alpha_K + n_K)
.
$$

 \State \(\triangleright\) \textbf{Step 3 (scale matrices):} For each \(k\), compute \(S_{(k)}\) and draw
$$
\Sigma_k \sim \mathcal{IW}(\nu_0 + n_k \nu_k,\;\Psi_0^{-1} + S_{(k)} ).
$$
\qquad If \(n_k=0\) draw from prior \(\mathcal{IW}(\nu_0,\Psi_0^{-1})\).

  \State \(\triangleright\) \textbf{Step 4 (degrees \(\nu_k\)):}
  \State  For each \(k\), propose \(\log\nu^\ast \sim 
  \mathcal{N} (\log\nu_k,\sigma_{\text{MH}}^2)\) and accept with probability
    \[
      \min \big\{1,\ \exp \big( \ell(\nu^\ast) + \log\nu^\ast - \ell(\nu_k)
      - \log\nu_k \big) \big\},
    \]
\qquad    where \(\ell(\nu)\) is the log posterior for cluster \(k\) computed using \(n_k,L_{(k)},S_{(k)}\) and \(\Sigma_k\).

  \State \(\triangleright\) \textbf{Record / monitor:} Compute log-likelihood; save samples.
\EndFor
\State \textbf{Return:} posterior draws of \(\bm\pi,\{\nu_k\},\{\Sigma_k\},\{z_i\}\) and monitoring diagnostics.
\end{algorithmic}
\end{algorithm}

{\bf Practical implementation:}
Computation of the trace terms is further optimized via the vectorization identity $\operatorname{tr}(A B) = \operatorname{vec}(A)^\top \operatorname{vec}(B)$, enabling simultaneous evaluation of $\operatorname{tr}(\Sigma_k^{-1} S_i)$ across all $i$ through the product of an $n \times p^2$ matrix, formed by rows of $\operatorname{vec}(S_i)^\top$, and the vector $\operatorname{vec}(\Sigma_k^{-1})$. 
Numerical precision is maintained by employing Cholesky decompositions for matrix inversions and determinant calculations—incorporating a minimal diagonal jitter (e.g., $10^{-6}I_p$) to condition ill-behaved matrices—and utilizing Bartlett decompositions for stable inverse-Wishart sampling. 
Finally, the algorithm adopts weakly informative priors (e.g., $\nu_0 = p+2$, $\Psi_0 = I_p$) and employs log-space MH proposals with scales $\sigma_{\text{MH}}^2$ (can be cluster-specific) tuned to acceptance rates of 20–40\%, utilizing log-sum-exp operations to ensure precision in density calculations.

\subsection{Bayesian MoE–Wishart model}

We put the following prior distributions on the parameters for the MoE–Wishart model in \eqref{eq_MoE_model}:  
\[
\Sigma_k\sim\mathcal{IW}(\nu_0,\Psi_0),
\quad
\nu_k\sim\text{Gamma}(a_\nu,b_\nu),
\quad
\beta_{1:(K-1)}\stackrel{\text{iid for columns}}{\sim}\mathcal N_q(0,\sigma_\beta^2 I_q).
\]
Let \(\Theta = \{z_{1:n},(\nu_k,\Sigma_k)_{k=1}^K,\beta_{1:(K-1)}\} \). 
The joint posterior is
\[
p(\Theta\mid S_{1:n},X_{1:n})
\propto
\prod_{i=1}^n \prod_{k=1}^K
\Big[
\pi_{ik}(X_i;\beta),
f_{\mathcal W}(S_i\mid\nu_k,\Sigma_k)
\Big]^{\mathbf 1_{\{z_i=k\}}}
\prod_{k=1}^K p(\Sigma_k)p(\nu_k)
\prod_{k=1}^{K-1} p(\beta_k).
\]
Conditioning on the latent labels yields a factorized posterior structure that enables efficient Gibbs sampling with a MH step for \(\nu_k\).

\noindent The entire Metropolis-within-Gibbs sampling for the MoE–Wishart model is as follows.

{\bf Step 1: Update labels $z_i$ (categorical sampling).}

The posterior probability of $z_i=k$ is
\[
\Pr(z_i=k\mid S_i,X_i,\Theta)\;=\;\frac{\exp(\ell_{ik})}{\sum_{\ell=1}^K\exp(\ell_{i\ell})},
\]
where $\Theta$ collects all coefficients $\beta_1$,..., $\beta_{K-1}$, and $z_i$ is sampled from this categorical distribution, with $ \ell_{ik} $ given in \eqref{eq_ell_ik}.

{\bf Step 2: Update gating coefficients $\beta_{1:(K-1)}$ (Metropolis--Hastings).}

We use MH sampling to update the coefficients in blocks by cluster.
Specifically, for each \(k=1,\ldots,K-1\), we propose
\[
\beta_k^\ast = \beta_k + \varepsilon_k,
\qquad
\varepsilon_k \sim \mathcal 
N \left(0,\sigma_{\beta_k}^2 I\right),
\]
while fixing \(\beta_K \equiv 0\) for identifiability. 
Let \(\pi_{ik}^\ast\) denote the gating probabilities computed under the proposed coefficients \(\beta^\ast\). 
Under a zero-mean Gaussian prior with variance \(\sigma_\beta^2\), the log prior difference is
$
\Delta \log p(\beta)
= -\frac{1}{2\sigma_\beta^2}
\sum_{k=1}^{K-1}
\big(
\|\beta_k^\ast\|_2^2 - \|\beta_k\|_2^2
\big).
$
Conditioned on the latent cluster labels \(z\), the gating likelihood is
$
\log p(\mathbf z \mid X,\beta)
= \sum_{i=1}^n \log \pi_{i,z_i}.
$
The MH log target ratio is therefore
\[
\log \alpha
=
\sum_{i=1}^n \big( \log \pi_{i,z_i}^\ast - \log \pi_{i,z_i} \big)
+
\Delta \log p(\beta).
\]
The proposal is accepted with probability \(\min \{1,\exp(\log \alpha) \} \).

{\bf Step 3: Update $\Sigma_k$ (conjugate inverse–Wishart).}

Each $\Sigma_k$ is updated sequentially from its inverse–Wishart posterior:
\[
\Sigma_k\mid\text{rest} 
\sim
\mathcal{IW}\!\Big(\nu_0 + n_k\nu_k,\; \Psi_0^{-1} + S_{(k)}\Big),
\]
where $n_k=|I_k|$, $I_k=\{i:z_i=k\}$, 
and $S_{(k)}=\sum_{i\in I_k}S_i$.

{\bf Step 4: Update degrees of freedom $\nu_k$ (MH   on log scale).}

Each $\nu_k$ is updated sequentially via a log-normal random-walk proposal:
\[
\log\nu_k^\ast\sim\mathcal{N}(\log\nu_k,\sigma_{\nu}^2),\qquad\nu_k^\ast=\exp(\log\nu_k^\ast).
\]
Using summarized quantities $S_{(k)}$ and $L_{(k)}=\sum_{i\in I_k}\log|S_i|$, the log joint density (data + prior) that depends on $\nu$ is
\[
\begin{aligned}
\ell(\nu) &= \sum_{i\in I_k}\Big\{\frac{\nu-p-1}{2}\log|S_i| - \frac{\nu p}{2}\log2 - \frac{\nu}{2}\log|\Sigma_k| - \log\Gamma_p\!\big(\tfrac{\nu}{2}\big)\Big\}\\
&\quad + (a_\nu-1)\log\nu - \nu b_\nu  + C,
\end{aligned}
\]
where $C$ collects terms not depending on $\nu$. Using $L_{(k)}$ and $\operatorname{tr}(\Sigma_k^{-1}S_{(k)})$ we compute differences efficiently. The MH log target ratio is
\[
\log\alpha = \big[\ell(\nu^\ast)+\log\nu^\ast\big] - \big[\ell(\nu)+\log\nu\big],
\]
where the extra $\log\nu$ terms are the change-of-variable Jacobian from the log proposal. Accept $\nu^\ast$ with probability $\min(1,\exp(\log\alpha))$.

\subsection{Selection of $K $ }

Let $K$ denote the number of experts in the mixture--of--experts (MoE) model defined in \eqref{eq_MoE_model}. Because increasing $K$ always weakly improves the in--sample likelihood, model selection must trade off goodness of fit against model complexity. We therefore select $K$ using information criteria and Bayesian predictive measures that are appropriate for latent--mixture models.

\subsubsection{Model dimension}

Let $p$ be the dimension of the Wishart random matrix $S_i$ and let $q$ denote the dimension of the covariate vector $X_i$ (including an intercept). 
Each expert $k$ contributes a Wishart scale matrix $\Sigma_k\in\mathbb S^p_{+}$ with $\tfrac{p(p+1)}2$ free parameters and a degrees--of--freedom parameter $\nu_k>p-1$. Hence, each expert contributes
$\frac{p(p+1)}2+1$
free parameters. The softmax gating network is identifiable only up to a baseline; setting $\beta_K=0$ yields $(K-1)q$ free gating parameters. The total dimension of the model is therefore
\[
\mathcal{D}_K = K\Big(\frac{p(p+1)}2+1\Big)+(K-1)q.
\]

\subsubsection{Bayesian Information Criterion}

Let $\hat\Theta_K$ denote the maximum likelihood or posterior mode estimator under a $K$--component model. The Bayesian Information Criterion (BIC) is defined as
\[
\mathrm{BIC}(K)
=
-2\ell(\hat\Theta_K)+ \mathcal{D}_K \log n,
\]
where $\ell(\Theta)$ is the observed--data log--likelihood in \eqref{eq_observe_log_likelihood_MoE}.

\subsubsection{Integrated Completed Likelihood}

Because mixture models may artificially increase $K$ by splitting existing components, we also consider the Integrated Completed Likelihood (ICL) \citep{biernacki2002assessing}, which penalizes posterior uncertainty in class assignments. 
Let $ r_{ik} = \Pr(z_i=k\mid S_i,X_i,\hat\Theta_K)$ denote the posterior responsibilities. The ICL criterion is
\[
\mathrm{ICL}(K)
=
\mathrm{BIC}(K)
-
2\sum_{i=1}^n\sum_{k=1}^K r_{ik}\log r_{ik}.
\]
The entropy term favors models with well--separated experts and thus selects the number of substantively distinct regimes.

\subsubsection{Bayesian LOO}

To compare Bayesian predictive models, out-of-sample pointwise predictive accuracy can be estimated using leave-one-out cross-validation (LOO) or the widely applicable information criterion (WAIC).
Both approaches are based on evaluating the log-likelihood at posterior draws, and WAIC is asymptotically equivalent to LOO.
However, \cite{Vehtari2017} show that LOO implemented via Pareto-smoothed importance sampling (PSIS) is more robust in finite samples, particularly in the presence of weak priors or influential observations.

The expected log pointwise predictive density (elpd) under LOO is defined as
\[
\text{elpd}_{\text{loo}} = \sum_{i=1}^n \log p(S_i \mid S_{-i}),
\qquad
p(S_i \mid S_{-i})
\approx
\left(
\frac{1}{N} \sum_{t=1}^N \frac{1}{p(S_i \mid \Theta^{(t)})}
\right)^{-1}.
\]
where \( S_{-i} \) denotes the data with the \( i \)-th observation removed, and $p(S_i \mid S_{-i})$ can be approximately estimated by the importance sampling using posterior draws \( \{\Theta^{(t)}\}_{t=1}^N \) from \( p(\Theta \mid S) \), where \( \Theta^{(t)} \) denotes the full parameter vector at the \( t \)-th MCMC iteration. 
Because the raw importance ratios can be heavy-tailed, PSIS replaces their upper tail with a fitted generalized Pareto distribution, yielding stabilized importance weights and a diagnostic shape parameter that indicates the reliability of the approximation. 
The resulting LOO estimator of predictive accuracy is
\[
\widehat{\text{elpd}}_{\text{loo}}
= \sum_{i=1}^n \widehat{\text{elpd}}_{\text{loo},i}
= \sum_{i=1}^n \log p(S_i \mid S_{-i}),
\]
where \( \widehat{\text{elpd}}_{\text{loo},i} \) denotes the contribution of observation \( i \).

\subsubsection{Model selection rule}

We select the number of experts $K$ by combining these complementary criteria. BIC provides consistency for large samples, ICL identifies the number of economically meaningful regimes, and elpd$_\text{loo}$ is particularly useful for assessing the predictive performance of Bayesian methods.
In practice, we choose the smallest $K$ that minimizes BIC and ICL, or maximizes elpd$_\text{loo}$ for Bayesian methods.

\section{Maximum likelihood and EM algorithm}
\label{sc_MLE_EM}

\subsection{Maximum likelihood estimation}
The inference of classical Wishart mixture models was carried out via the Expectation–Maximization (EM) algorithm in \cite{Hidot2010}. 
Here we estimate the parameters of the mixture-of-experts (MoE) model in \eqref{eq_MoE_model} by maximum likelihood using the EM algorithm. The EM framework is particularly well suited here because the likelihood involves a latent mixture structure induced by the unobserved component labels.

Let \(\Theta = \{(\beta_k,\nu_k,\Sigma_k)_{k=1}^K \} \) denote the model parameters under the MoE formulation. 
The observed-data log-likelihood of the MoE model \eqref{eq_MoE_model} is:
\begin{equation}
\label{eq_observe_log_likelihood_MoE}
    \ell(\Theta) 
    =
    \sum_{i=1}^n \log \left( \sum_{k=1}^K \pi_{ik}(X_i;\beta)
    \cdot
    f_{\mathcal W}(S_i \mid \nu_k,\Sigma_k) \right)
.
\end{equation}
Because of the summation inside the logarithm, direct maximization is analytically intractable. The EM algorithm overcomes this difficulty by introducing the latent class labels $z_i$ and iteratively maximizing a tractable surrogate objective.

If the latent cluster labels $z_i$ were observed, the complete-data log-likelihood (up to additive constants) would be:
$$
\ell_c(\Theta) = \sum_{i=1}^n \sum_{k=1}^K z_{ik} \Big[ \log \pi_{ik}(X_i;\beta) + \log f_{\mathcal W}(S_i \mid \nu_k,\Sigma_k) \Big],
$$
which, after inserting the Wishart density, becomes:
\begin{multline}
  \ell_c(\Theta) 
= 
\sum_{i=1}^n \sum_{k=1}^K z_{ik} 
\left[ \frac{\nu_k-p-1}{2}\log|S_i| -\frac{\nu_k}{2}\log|\Sigma_k| -\frac12 \operatorname{tr}(\Sigma_k^{-1}S_i) 
\right.
\\
\left.
-\log\Gamma_p\left(\frac{\nu_k}{2}\right) -\frac{\nu_k p}{2}\log 2 + \log \pi_{ik}(X_i;\beta) \right]
.  
\end{multline}
This representation reveals that the gating parameters $\beta$ and the expert parameters $(\nu_k, \Sigma_k)$ enter the likelihood in separable blocks, which yields convenient conditional maximization steps.

\subsection{EM algorithm}

Let $z_i \in \{1,\dots,K\}$ denote the latent cluster labels of observation $i$, and define the posterior responsibility:
$$
r_{ik} := \Pr(z_i = k \mid S_i, X_i, \Theta).
$$
We introduce some summarized quantities to simplify notation:
$$
n_k = \sum_{i=1}^n r_{ik}, \qquad M_k = \sum_{i=1}^n r_{ik} S_i, \qquad \overline{\log |S|}_k = \frac{1}{n_k} \sum_{i=1}^n r_{ik} \log |S_i|, \qquad \overline S_k = \frac{M_k}{n_k}.
$$

The EM algorithm alternates between computing ``soft'' assignments (responsibilities) of observations to experts via the gating network (E-step) and updating the gating coefficients $\beta$ via weighted multinomial logistic regression and the expert parameters $(\nu_k,\Sigma_k)$ via weighted maximum likelihood (M-step). This alternating optimization usually yields a monotone non-decreasing observed-data log-likelihood at every iteration, and provides a stable, statistically principled estimator for the MoE–Wishart model.

\noindent {\bf E-step.}
At iteration $(t)$, we compute the posterior responsibilities:
$$
r_{ik}^{(t)} 
=
\Pr(z_i=k \mid S_i, X_i, \Theta^{(t)}) 
=
\frac{\pi_{ik}(X_i;\beta^{(t)}) 
f_{\mathcal W}
(S_i\mid \nu_k^{(t)}, \Sigma_k^{(t)})}{\sum_{j=1}^K \pi_{ij}(X_i;\beta^{(t)}) f_{\mathcal W}(S_i\mid \nu_j^{(t)}, \Sigma_j^{(t)})}.
$$
These weights quantify the probability that observation $i$ originates from expert $k$, given the current parameter estimates.

\noindent {\bf M-step.}
Maximizing the conditional expectation of the complete-data log-likelihood:
$$
Q(\Theta \mid \Theta^{(t)}) = \sum_{i=1}^n \sum_{k=1}^K r_{ik}^{(t)} \Big[ \log \pi_{ik}(X_i;\beta) + \log f_{\mathcal W}(S_i \mid \nu,\Sigma) \Big].
$$

$\bullet$ First, update of gating network ($\beta$):
The update for $\beta$ corresponds to a weighted multinomial logistic regression:
$$
\beta^{(t+1)} 
=
\arg\max_\beta \sum_{i=1}^n \sum_{k=1}^K r_{ik}^{(t)} \log \pi_{ik}(X_i;\beta).
$$
This optimization problem has no closed-form solution and is solved numerically, for example via BFGS or Newton–Raphson method.

$\bullet$ Second, update of scale matrices ($\Sigma_k$):
For fixed $\nu_k$, the maximizer of $Q(\Theta \mid \Theta^{(t)})$ with respect to $\Sigma_k$ admits a closed-form expression:
$$
\Sigma_k^{(t+1)} = \frac{1}{n_k \nu_k^{(t+1)}} \sum_{i=1}^n r_{ik}^{(t)} S_i = \frac{\overline S_k}{\nu_k^{(t+1)}}.
$$

$\bullet$ Third, update of degrees of freedom ($\nu_k$):
The update for $\nu_k$ is obtained by solving the score equation:
$$
\psi_p\left(\frac{\nu_k}{2}\right) - \log \nu_k + \log\left(\frac{|M_k|}{n_k}\right) - \overline{\log|S|}_k + p\log 2 - p\log n_k = 0,
$$
where $\psi_p(\cdot)$ denotes the multivariate digamma function. This nonlinear equation is solved numerically using Newton–Raphson or a similar root-finding method.

\section{Simulations studies}
\label{sc_simulations}

The code to reproduce the simulation results is available at \url{https://github.com/zhizuio/moewishart/tree/main/numerical_studies}.

\subsection{Simulation with mixture model}
\subsubsection{Setup}\label{simSetup1}

We consider a finite mixture of Wishart distributions with \(K=3\) clusters. The mixture weights are fixed at
$ \;
\bm{\pi} = (\pi_1,\pi_2,\pi_3) = (0.35,0.40,0.25) \; .
$
We vary sample sizes \(n \in \{200, 500, 1000\}\) and matrix dimensions 
\(\; p \in \{2, 8\}\).
Each positive definite observed covariance matrix \(S_i\), \(i=1,\ldots,n\), is  generated independently from the mixture model
\[
S_i \sim \sum_{k=1}^3 \pi_k  \mathcal{W}_p(\nu_k, \Sigma_k),
\]
where \(\mathcal{W}_p(\nu_k,\Sigma_k)\) denotes the \(p\)-dimensional Wishart distribution with degrees of freedom \(\nu_k\) and scale matrix \(\Sigma_k\).
The degrees of freedom are specified with two cases corresponding to the two options of the matrix dimension $p$:
\[
(\nu_1,\nu_2,\nu_3)=
\begin{cases}
(8, 12 ,3), & \text{if } p=2,
\\
(9, 20 ,14), & \text{if } p=8.
\end{cases}
\]
Similarly the scale matrices $\Sigma_k$'s are  specified with cases corresponding to the matrix dimensions:
\begin{itemize} 
\item If $p=2$, we set $\Sigma_1 =
\begin{pmatrix}
  0.5 & 0.2\\ 
  0.2 & 0.7
\end{pmatrix}$,
$\Sigma_2 =
\begin{pmatrix}
  2.0 & 0.6\\ 
  0.6 & 1.5
\end{pmatrix}$,
$\Sigma_3 =
\begin{pmatrix}
  4.0 & 0.2\\ 
  0.2 & 3.0
\end{pmatrix}$.
\item If $p=8$, we take 
$\Sigma_1 = \{0.5^{|j-j'|}\}_{jj'}$, 
$\Sigma_2 = \{0.2^{|j-j'|}\}_{jj'}$, 
$\Sigma_3 = \{0.8^{|j-j'|}\}_{jj'}$, 
$j,j'\in\{1,...,p\}$.
\end{itemize}

\noindent {\bf Considered methods:}
For each simulation setup, we evaluate four competing estimation approaches:
\begin{itemize}
    \item Two are based on finite mixture models: the full Bayesian method (denoted by \textbf{Bayes}) and the maximum likelihood estimator implemented via the EM algorithm (denoted by \textbf{EM}). 

    \item The other two approaches are based on the mixture-of-experts framework: the full Bayesian approach (denoted by \textbf{Bayes-MoE}) and the EM-based maximum likelihood (denoted by \textbf{EM-MoE}). More specifically, only intercepts are included for both \textbf{Bayes-MoE} and \textbf{EM-MoE} as concomitant covariates and the gating probabilities $\pi_{ik}$ are reduced to $\pi_k$.
\end{itemize}
For the two full Bayesian versions (i.e., \textbf{Bayes} and \textbf{Bayes-MoE}), we run MCMC with $20000$ iterations, of which the first $5000$ iterations are discarded as the burn-in period.

\noindent {\bf Model performance evaluation:}
we  use the average over all components errors to access the performance of the considered methods. More specifically, we consider the following metrics: 
$$
\frac{1}{K}\|\hat{\bm\pi}-\bm\pi\|_1
; \quad
\frac{1}{K}\|\hat{\bm\nu}-\bm\nu\|_1
; \quad
\frac{1}{K}\sum_{k=1}^K\|\hat{\Sigma}_k-\Sigma_k\|_2^2
.
$$
For the two full Bayesian versions (i.e., \textbf{Bayes} and \textbf{Bayes-MoE}), posterior mean estimators are used to calculate the evaluation metrics. 
We generate $100$ data sets for each simulation setting and report the results in Figure \ref{fig:simMM}.

\subsubsection{Simulations results}

\begin{figure}[!htp]

    \centering
    \includegraphics[width=16cm]{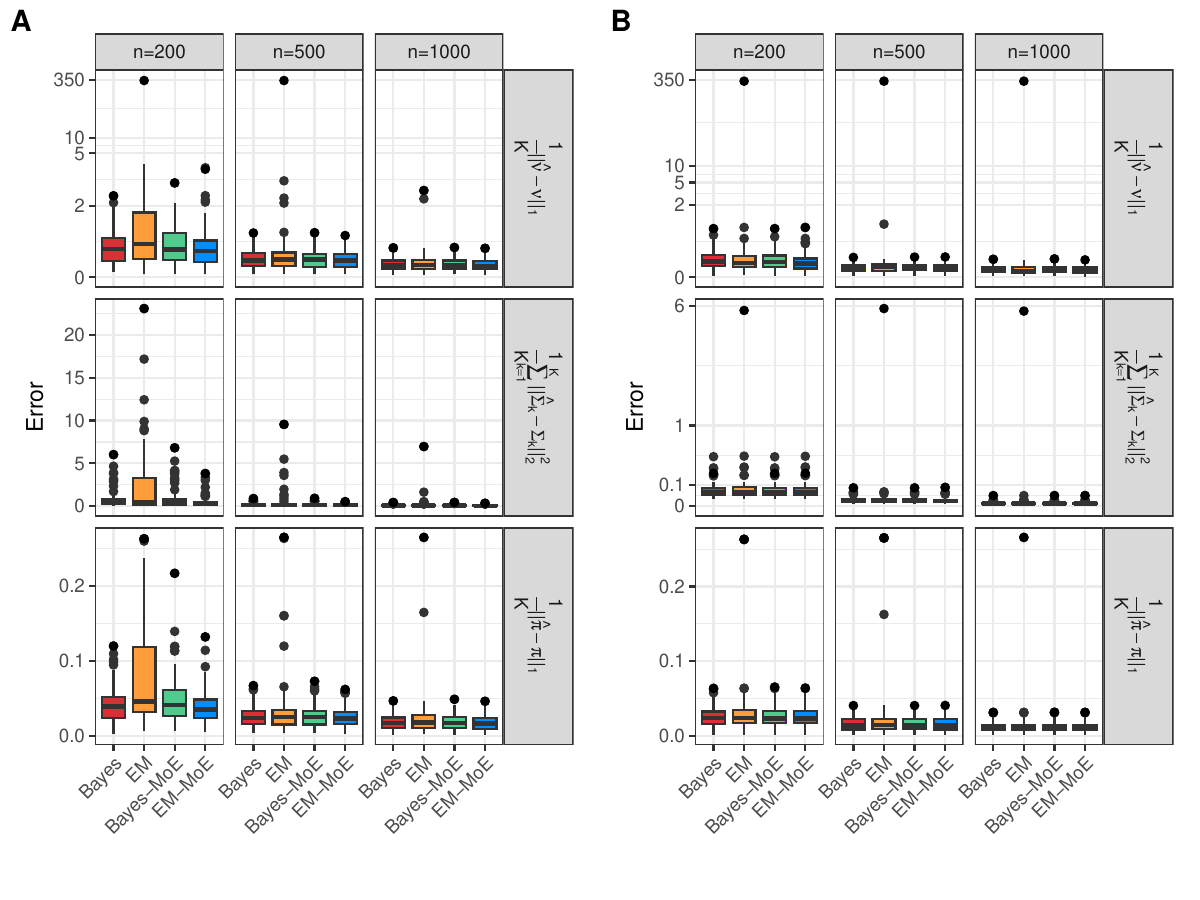}
    \caption{Simulation results. Data were generated from finite mixture models over 100 Monte Carlo replications. Results are reported for sample sizes $n \in \{200, 500, 1000\} $ under two dimensional settings: (A) $p=2$ and (B) $p=8$. The four competing methods compared include: (i) Bayesian mixture models (\textbf{Bayes}); (ii) EM-based mixture models (\textbf{EM}); (iii) Bayesian mixture-of-experts models (\textbf{Bayes-MoE}); and (iv) EM-based mixture-of-experts models (\textbf{EM-MoE}). Estimation performance is evaluated using average componentwise errors:
$
\frac{1}{K}\|\hat{\bm\pi}-\bm\pi\|_1
; \;
\frac{1}{K}\|\hat{\bm\nu}-\bm\nu\|_1
; \;
\frac{1}{K}\sum_{k=1}^K\|\hat{\Sigma}_k-\Sigma_k\|_2^2
. $ 
}
        \label{fig:simMM}
\end{figure}

Figure \ref{fig:simMM} illustrates that the estimation accuracy of all four methods—\textbf{Bayes}, \textbf{EM}, \textbf{Bayes-MoE}, and \textbf{EM-MoE}—for the parameters $\boldsymbol{\nu}$, $\Sigma_k$, and $\boldsymbol{\pi}$ improves systematically with increasing sample size under data generated from the mixture model.

For $p=2$, Figure \ref{fig:simMM}A shows that the correctly specified mixture-model \textbf{EM} approach yields larger estimation errors than the other three approaches at the smallest sample size ($n=200$) and large outliers at sample sizes $n\in\{500,1000\}$. 
These outliers arise because the estimates $\bm\nu$, $\Sigma_{1:K}$ and $\bm\pi$ obtained by \textbf{EM} failed to converge for those simulated data sets.
The correctly specified mixture-model \textbf{Bayes} and mixture-of-experts methods (\textbf{Bayes-MoE} and \textbf{EM-MoE}) achieve similar estimation errors for $\boldsymbol{\nu}$, $\Sigma_{1:K}$, and $\boldsymbol{\pi}$ across all sample sizes considered ($n \in {200, 500, 1000}$). 
Figure \ref{fig:simMM}B for $p=8$ indicates that all four approaches have comparable performance, except that the \textbf{EM} approach consistently produces a few outliers due to convergence issues of $\bm\nu$, $\Sigma_{1:K}$ and $\bm\pi$ for some simulated data sets.

Supplementary Figures \ref{figS:simMM} ($p=2$) and \ref{figS:simMM_p8} ($p=8$) demonstrate stable Markov chain Monte Carlo (MCMC) behavior for the Bayesian methods (\textbf{Bayes} and \textbf{Bayes-MoE}), as evidenced by the log-likelihood trace plots for each working model. In addition, cluster-specific parameters—$\pi_k$, $\nu_k$, and $\log|\Sigma_k|$ for \textbf{Bayes}, and $\nu_k$ and $\log|\Sigma_k|$ for \textbf{Bayes-MoE} ($k \in {1,2,3}$)—exhibit satisfactory mixing. As expected, larger fluctuations in the trace plots are observed for $n=200$ relative to $n=500$ and $n=1000$. 
Table \ref{tab:simMM_ESS} reports effective sample sizes (ESS) for selected parameters, showing systematically smaller ESS values with larger variations at $n=200$ compared to larger sample sizes for both Bayesian approaches. 
Finally, Supplementary Figures \ref{figS:simEM}A–B confirm that the EM algorithms, for both the mixture and MoE working models, converge reliably in terms of the log-likelihood across all simulation scenarios based on one simulated data set.

\begin{table}[!htp]
\centering
\caption{Effective sample sizes (ESS) from MCMC for Bayes and Bayes-MoE methods.
Data is generated from the mixture model. Reported are the mean and standard deviation of ESS over 100 simulations for selected parameters.
\label{tab:simMM_ESS} }
  \small
\begin{tabular}{r cc c cc c cc}  
\toprule
\multicolumn{1}{l}{\textbf{Working model}} 
  & \multicolumn{2}{c}{$\bm\nu$} 
  && \multicolumn{2}{c}{$\Sigma$} 
  && \multicolumn{2}{c}{$\bm\pi$} 
  \\
\cmidrule{2-3}
\cmidrule{5-6} 
\cmidrule{8-9} 
 & $\nu_1$ & $\nu_2$ &
 & $\Sigma_{1,11}$ & $\Sigma_{2,11}$ &
 & $\pi_1$ & $\pi_2$  
 \\
\midrule 
\multicolumn{1}{l}{\textbf{Mixture model}: $p=2$ } 
\medskip\\
$n=200$ &
 675 (250) & 164 (64) && 1079 (525) & 214 (88) && 395 (244) & 907 (865)
\\ 
$n=500$ & 
825 (140) & 204 (43) && 1339 (349) & 259 (61) && 500 (155) & 1040 (455)
\\
$n=1000$ & 
794 (110) & 183 (31) && 1192 (298) & 222 (45) && 433 (107) & 894 (331)
\bigskip\\
\multicolumn{1}{l}{\textbf{Mixture model}: $p=8$ } 
\medskip\\
$n=200$ & 
1295 (91) & 760 (69) && 8528 (797) & 3303 (514) && 14651 (481) & 14641 (541)
\\ 
$n=500$ & 
1413 (85) & 838 (47) && 8907 (604) & 3709 (338) && 14922 (323) & 14965 (263)
\\
$n=1000$ & 
1340 (87) & 829 (56) && 9179 (651) & 3559 (324) && 14990 (100) & 14992 (99)
\\
\\
\midrule 
\multicolumn{2}{l}{\textbf{MoE model}: $p=2$ } 
\medskip\\
$n=200$ & 
622 (266) & 170 (70) && 1029 (521) & 228 (227) && - & - 
\\
$n=500$ &     
 713 (158) & 179 (36) && 1085 (322) & 213 (47) && - & - 
\\
$n=1000$ &     
678 (138) & 176 (33) && 999 (267) & 199 (44) && - & - 
\bigskip\\
\multicolumn{2}{l}{\textbf{MoE model}: $p=8$ } 
\medskip\\
$n=200$ &  
1315 (91) & 814 (66) && 8784 (871) & 3597 (462) && - & - 
\\
$n=500$ &  
1387 (73) & 851 (49) && 9069 (570) & 3610 (301) && - & - 
\\
$n=1000$ &  
1349 (54) & 821 (47) && 8837 (607) & 3565 (316) && - & - 
\medskip\\
\bottomrule
\end{tabular}
\end{table}

\begin{figure}[!htp]
    \centering
    \includegraphics[width=15cm]{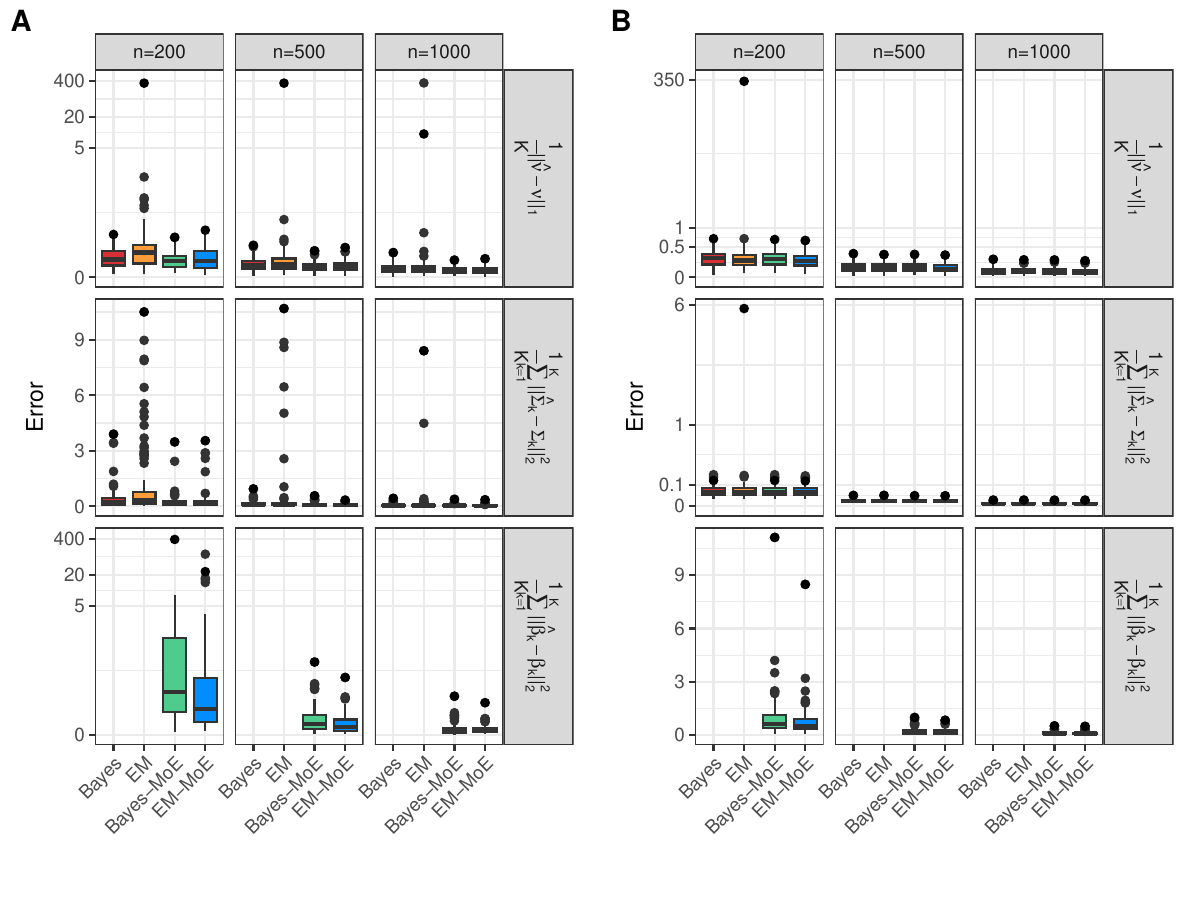}
    \caption{Simulation results. Data were generated from the \textbf{mixture-of-experts (MoE) model} across 100 simulated data sets. Panels (A) and (B) correspond to the settings $p=2$ and $p=8$, respectively. The four competing methods compared are: (i) Bayesian inference under the mixture model (\textbf{Bayes}); (ii) EM-based inference under the mixture model (\textbf{EM}); (iii) Bayesian inference under the mixture-of-experts model (\textbf{Bayes-MoE}); and (iv) EM-based inference under the mixture-of-experts model (\textbf{EM-MoE}). 
    }
    \label{fig:simMoE}
\end{figure}

\subsection{Simulation with mixture–of–experts model}
\subsubsection{Setup}
We extend the finite mixture of Wishart distributions to a mixture–of–experts setting with covariate-dependent mixing proportions. 
As before, we take $K=3$ latent clusters, and examine sample sizes $n \in \{200, 500, 1000\}$ and matrix dimensions $p \in \{2, 8\}$.

With $q=3$, 
let $X \in \mathbb{R}^{n \times q}$ denote the covariate matrix
where the entries are independently generated as $x_{ij}\sim \mathcal{N}(0,1)$. 
The mixing proportions are obtained via a multinomial logistic (softmax) function
$$
\pi_{ik} ( X_i; \beta )
\;=\; 
\frac{\exp\{X_i^\top \beta_k\}}
{\sum_{l=1}^K \exp\{X_i^\top \beta_l\}},
\quad i=1,\ldots,n,\; k=1,\ldots,K,
$$
with covariate effects $\beta_{1:K} \in \mathbb{R}^{q\times K}$ and the reference constraint $\beta_K=\mathbf{0}$ for identifiability. 
Each coordinate of the effect $ \beta $ is independently simulated from a uniform distribution, 
$\beta_{jk}\sim\mathcal U(-2,2)$, and $\beta_{1:K}$ is the same for every simulation.

Conditional on $X_i$, each positive definite matrix $S_i$ is independently generated from the mixture–of–experts model
$$
S_i \sim \sum_{k=1}^3
\pi_{ik}  ( X^\top \beta )
\,
\mathcal{W}_p(\nu_k, \Sigma_k),
\quad i=1,\ldots,n
$$
Here the cluster parameters $(\nu_k,\Sigma_k)$ are the same as in the finite mixture setup in Section \ref{simSetup1}.

\noindent {\bf Model performance evaluation:}
we  use the average over all cluster errors to access the performance of the considered methods. More specifically, we consider the following metrics: 
$$
\frac{1}{K}\|\hat{\bm\nu}-\bm\nu\|_1
; \quad
\frac{1}{K}\sum_{k=1}^K\|\hat{\Sigma}_k-\Sigma_k\|_2^2
; \quad
\frac{1}{K}\sum_{k=1}^K\|\hat{\beta}_k-\beta_k\|_2^2
.
$$
All other setups remain the same as the finite mixture case in Section \ref{simSetup1}.

\subsubsection{Simulations results}

As in the mixture-model setting, Figure \ref{fig:simMoE} demonstrates that estimation accuracy under the mixture-of-experts model improves monotonically with increasing sample size (Figure \ref{fig:simMoE}A for $p=2$ and Figure \ref{fig:simMoE}B for $p=8$). 
\textbf{Bayes-MoE} and \textbf{EM-MoE} exhibit comparable performance across all scenarios. 
At the smallest sample size ($n=200$) for both $p=2$ and $p=8$, \textbf{Bayes-MoE} yields slightly smaller estimation error for $\boldsymbol{\nu}$ than \textbf{EM-MoE}, while incurring marginally larger error for $\boldsymbol{\beta}$ (bottom-left panels).
The mixture-model-based methods (\textbf{Bayes} and \textbf{EM}) perform slightly worse than the correctly specified MoE-based approaches, with \textbf{EM} consistently producing the largest estimation errors across metrics for $p=2$. 
For $p=8$, \textbf{EM} also consistently produces a few outliers due to convergence issues of $\bm\nu$ and $\Sigma_{1:K}$ for some simulated data sets.

For data generated from the MoE model, Supplementary Figures \ref{figS:simMoE} ($p=2$) and \ref{figS:simMoE_p8} ($p=8$) further indicate stable Markov chain Monte Carlo (MCMC) behavior for both Bayesian methods, as evidenced by log-likelihood trace plots and component-specific parameter trajectories for each working model. Consistent with the mixture-model results, larger fluctuations are observed at $n=200$ relative to $n=500$ and $n=1000$. Table \ref{tab:simMoE_ESS} reports effective sample sizes (ESS) for selected parameters, which increase systematically with sample size for both Bayesian approaches. Finally, Supplementary Figure \ref{figS:simEM}C-D confirms convergence of the EM algorithms in terms of the log-likelihood for both the mixture and mixture-of-experts working models across both scenarios $p\in\{2,8\}$.

\subsection{Results on selecting the true cluster $ K $ }

We compared information-criterion-based model selection of the number of clusters for mixture and MoE models using the criteria introduced in Section 4.3: elpd$_\text{loo}$ and ICL for Bayesian working models, and BIC and ICL for EM working models. 
Under simulated settings with true $K=3$ (for data generated from either the mixture model or MoE model), we observed: 
(i) Bayesian MoE model (\textbf{Bayes-MoE}): elpd$_\text{loo}$ correctly identified $K=3$ (Table \ref{tab:sim_selectK_n200}). 
(ii) Bayesian mixture model (\textbf{Bayes}): elpd$_\text{loo}$ shows a mild tendency to favor larger $K$ than the truth. 
(iii) EM working models (\textbf{EM} and \textbf{EM-MoE}): BIC correctly identified $K=3$. 
Across all settings, ICL was systematically more conservative than the other criteria and preferred smaller $K$ (i.e., slightly under-splitting relative to the truth). 

For higher-dimensional ($p=8$) the mixture and MoE models with true $K=3$, Supplementary Table \ref{tabS:sim_selectK_n1000_p8} indicates that elpd$_\text{loo}$ shows a mild tendency to favor larger $K$ than the truth for Bayesian working models consistent with the $p=2$ results, whereas ICL correctly identified $K=3$ in all scenarios. Similar to the $p=2$ case, both BIC and ICL from the EM working models (\textbf{EM} and \textbf{EM-MoE}) correctly identified $K=3$ across all scenarios, except that \textbf{EM} selects $K=4$ for data generated from the MoE model.

\begin{table}[!htp]
\centering
\caption{Selection of the number of clusters on simulated data generated by the mixture and MoE models with $n=200$, $p=2$ and true $K=3$. Reported are the mean (standard deviation) of each information criterion over 100 simulations. Bayesian working models are evaluated with elpd$_\text{loo}$ (larger is better) and ICL (smaller is better); EM working models are evaluated with BIC and ICL (both smaller is better). Optimal values per method are bolded. 
\label{tab:sim_selectK_n200} }
\begin{tabular}{r ccccc}  
\toprule
\multicolumn{1}{c}{$K$} 
 & 2 & 3 & 4 & 5 & 6 
  \\
\midrule 
\multicolumn{6}{c}{\textbf{Data generated by mixture model}}
  \\
\midrule 
\multicolumn{1}{l}{\textbf{Bayes}} 
\\
elpd$_\text{loo}$ &  
-802.9 (33.5) & -776.2 (33.5) & -775.7 (33.6) & -775.1 (33.5) & \textbf{-774.6} (33.5)
\\
ICL & 
\textbf{1692.0} (70.3) & 1753.3 (72.5) & 1915.9 (86.5) & 2033.3 (85.5) & 2143.6 (93.7)
\smallskip\\
\multicolumn{1}{l}{\textbf{Bayes-MoE}} 
\\
elpd$_\text{loo}$ &  
-1863.5 (33.4) & \textbf{-1838.6} (33.5) & -1839.0 (33.5) & -1839.0 (33.5) & -1839.2 (33.6)
\\
ICL & 
\textbf{4030.0} (63.1) & 4158.8 (68.0) & 4241.9 (79.6) & 4303.5 (85.6) & 4383.7 (89.9)
\smallskip\\
\multicolumn{1}{l}{\textbf{EM}} 
\\
BIC &  
3751.5 (66.9) & \textbf{3725.1} (69.4) & 3736.1 (68.2) & 3749.7 (68.6) & 3766.3 (68.1)
\\ 
ICL &  
\textbf{3796.1} (70.5) & 3829.1 (76.2) & 3857.9 (73.5) & 3869.1 (75.4) & 3889.4 (86.1)
\\
\multicolumn{1}{l}{\textbf{EM-MoE}} 
\\
BIC &  
3751.9 (67.8) & \textbf{3720.5} (67.3) & 3737.7 (67.9) & 3755.3 (67.1) & 3772.3 (67.1)
\\ 
ICL &  
\textbf{3797.0} (72.6) & 3848.6 (71.8) & 3894.5 (80.8) & 3932.9 (77.4) & 3958.7 (80.3)
\\ 
\\
\multicolumn{6}{c}{\textbf{Data generated by MoE model} }
\\
\midrule 
\multicolumn{1}{l}{\textbf{Bayes}} 
\\
elpd$_\text{loo}$ &  
-830.0 (30.7) & -804.3 (29.5) & -803.8 (29.5) & -803.6 (29.5) & \textbf{-803.2} (29.5)
\\
ICL & 
\textbf{1774.4} (67.6) & 1851.7 (62.0) & 2024 (71.7) & 2152.1 (73.5) & 2253.5 (76.0)
\smallskip\\
\multicolumn{1}{l}{\textbf{Bayes-MoE}} 
\\
elpd$_\text{loo}$ &  
-1850.5 (31.2) & \textbf{-1812.9} (31.2) & -1813.8 (31.3) & -1814.1 (31.5) & -1815.1 (31.5)
\\
ICL & 
\textbf{3902.1} (69.2) & 3960.0 (68.7) & 4008.5 (70.3) & 4061.1 (71.8) & 4110.9 (70.2)
\smallskip\\
\multicolumn{1}{l}{\textbf{EM}} 
\\ 
BIC &  
3821.2 (60.5) & \textbf{3812.0} (59.9) & 3840.7 (58.5) & 3874.0 (59.6) & 3904.1 (59.3)
\\ 
ICL &  
\textbf{3876.0} (66.3) & 3939.0 (63.2) & 3979.9 (66.3) & 4015.3 (75.6) & 4047.5 (78.8)
\smallskip\\
\multicolumn{1}{l}{\textbf{EM-MoE}} 
\\ 
BIC &  
3734.5 (61.4) & \textbf{3688.2} (62.4) & 3714.9 (62.1) & 3741.7 (63.4) & 3768.3 (63.7)
\\ 
ICL &  
\textbf{3769.8} (65.6) & 3775.3 (64.7) & 3822.6 (73.3) & 3861.6 (74.6) & 3888.4 (77.2)
\\ 
\bottomrule
\end{tabular}
\end{table}

\clearpage
\section{Application to Cancer drug screening data}
\label{sc_application_drug}

\subsection{Data processing}
\begin{figure}[!ht]
    \centering
    \includegraphics[width=12cm]{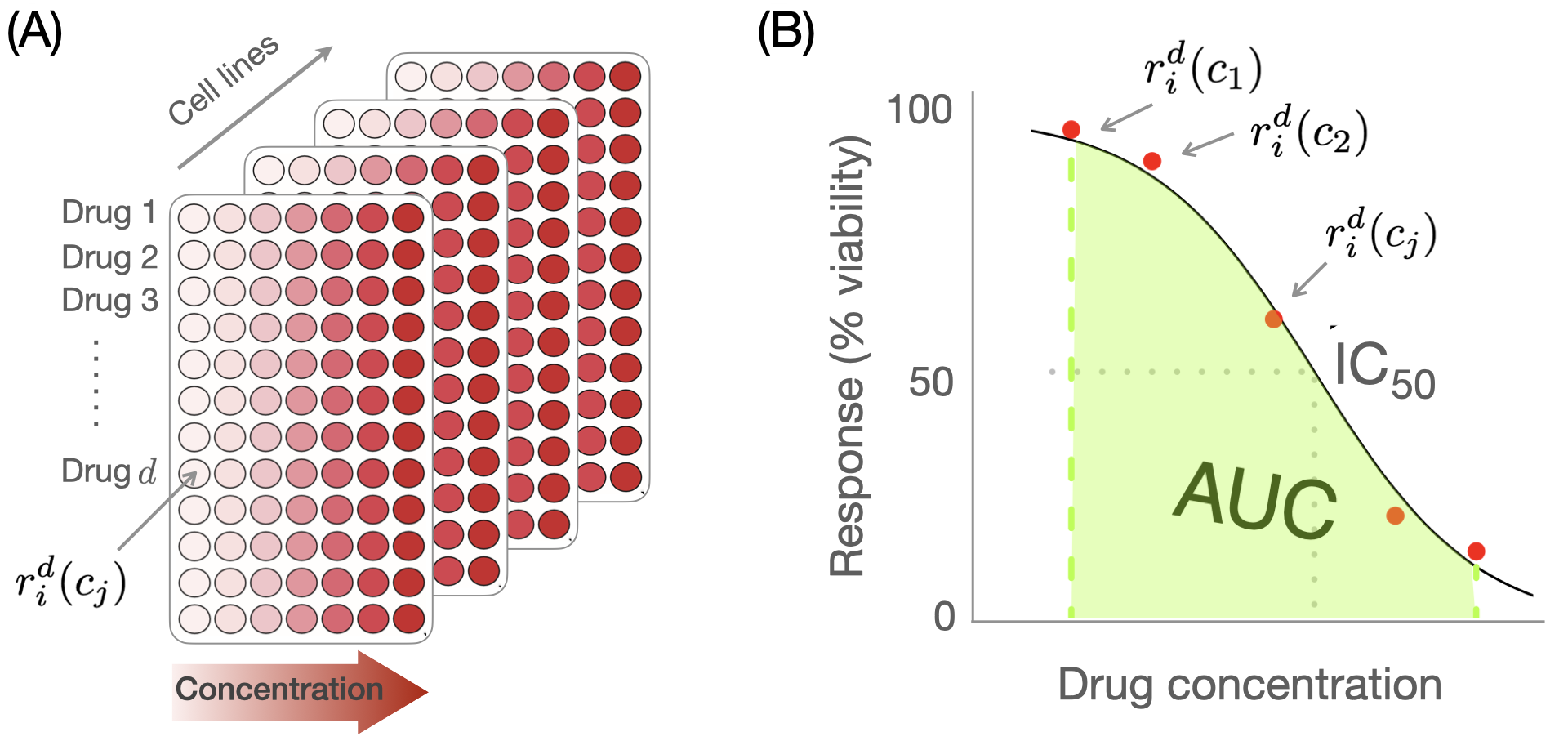}
    \caption{Experimental setup and data for cell lines from patient biopsies treated with drugs at multiple concentrations. (A) Experiment setup for an individual cell line experiment in multi-well cell culture plates. (B) Drug dose/concentration–response relationship for data from one row of wells of the plate. AUC is the area under the dose–response curve, often obtained from a fitted sigmoid function. IC$_{50}$ (half-maximal inhibitory concentration) represents the concentration of a drug required to reduce cell viability (survival) by $50\%$ compared to a control.}
    \label{fig:Drugsensitvity}
\end{figure}

The Cancer Therapeutics Response Portal (CTRP) v2 is a large-scale cancer cell line drug screening resource comprising responses of 481 compounds across 860 cell lines representing 24 primary tumor types \citep{SeashoreLudlow2015}.
The data are publicly available at \url{https://portals.broadinstitute.org/ctrp}. 
Let $r_i^d(c_j)$ denote the viability (or inhibition) of cell line $i$ ($i=1,...,860$) 
at drug concentration/dose $c_j$ with  $j=1,...,p$ for drug $d$, where $d=1,...,481$.  
For a given drug $d$, let $n_d$ be the number of cell lines with measured responses. 
Let 
$\bm r_i^d = (r_i^d(c_1),...,r_i^d(c_p))^\top \in \mathbb R^p$, and 
$\bar{\bm r}^d = (\bar r^d(c_1),...,\bar r^d(c_p))^\top \in \mathbb R^p$, 
where $\bar r^d(c_j)$ is the average response at dose $c_j$ across the $n_d$ cell lines with measurements for drug $d$. 
The empirical per-drug dose-dose covariance is 
$$
 S^{(d)} = \frac{1}{n_d -1}\sum_{i=1}^{n_d}  (\bm r_i^d - \bar{\bm r}^d)(\bm r_i^d - \bar{\bm r}^d)^\top \in \mathbb R^{p\times p}.
$$
This dose–response representation leverages the full dose–response information without reducing it to a single-parameter summary (e.g., IC50 or area under the dose–response curve (AUC); Figure \ref{fig:Drugsensitvity}) and avoids imposing a specific sigmoid form when curve fitting is unreliable. 
By modeling covariance across drug dosages, $S^{(d)}$ captures how inhibition at low, intermediate and high doses co-varies across cell lines for each drug, thereby preserving shapes along the dose axis.

To permit within-drug dose–dose covariance estimation, we analyzed a subset of $n=374$ drugs profiled at the same $p=5$ concentrations: 0.002, 0.016, 0.130, 1.000 and 8.300{\textmu}M. 
Using replicate viability measurements, we computed one $p\times p$ covariance matrix $S^{(d)}$ per drug, yielding $n=374$ matrices.

We model the per-drug dose–dose covariance matrices using finite mixture and mixture-of-experts models to differentiate latent drug classes. 
Drugs with similar mechanism-of-action (MoA) are expected to induce similar dose–response shape relationships across cell lines (e.g., how low-, mid- and high-dose effects co-vary); such patterns are encoded in the covariance matrices $S^{(d)}$.

The MoE model formulation additionally accommodates drug- or context-specific covariates (molecular structures and/or genetic information) to sharpen class separation and aid interpretation with respect to molecular targets.  
For the MoE models, we incorporated two drug-level covariates. 
The first covariate is drug status, coded as a binary indicator (0: approved by the FDA or used in clinical trials; 1: experimental compound). 
The second covariate is derived from the compound SMILES (Simplified Molecular Input Line Entry System), which encodes chemical structure. 
SMILES strings were converted to molecular fingerprints; we then performed principal component analysis on the fingerprint matrix and used the first principal component as a covariate in the MoE models.

\subsection{Results}
Model selection results are summarized in Supplementary Table \ref{tabS:realDrug_selection}. 
For the Bayesian mixture model (\textbf{Bayes}), $K=5$ and $K=7$ yield the highest elpd$_\text{loo}$ values, with $K=5$ and $K=7$ essentially tied. 
The MCMC diagnostics for $K=5$ (Supplementary Figure \ref{figS:realMM}) indicate stable sampling behavior, and $K=5$ is also preferred by ICL. 
Consistent with these findings, the Bayesian MoE model (\textbf{Bayes-MoE}) achieves its best elpd$_\text{loo}$ at $K=6$, but $K=5$ yields comparable predictive performance (Supplementary Table \ref{tabS:realDrug_selection}), stable MCMC behavior (Supplementary Figure \ref{figS:realMoE}), and is favored by ICL. 
We therefore use $K=5$ for both Bayesian methods in subsequent analyses.  
In contrast, the EM-based mixture and MoE models selected $K=10$ and $K=11$, respectively, under both BIC/ICL (see Supplementary Table \ref{tabS:realDrug_selection}).

\begin{figure}[!htp]
    \centering
    \includegraphics[width=15cm]{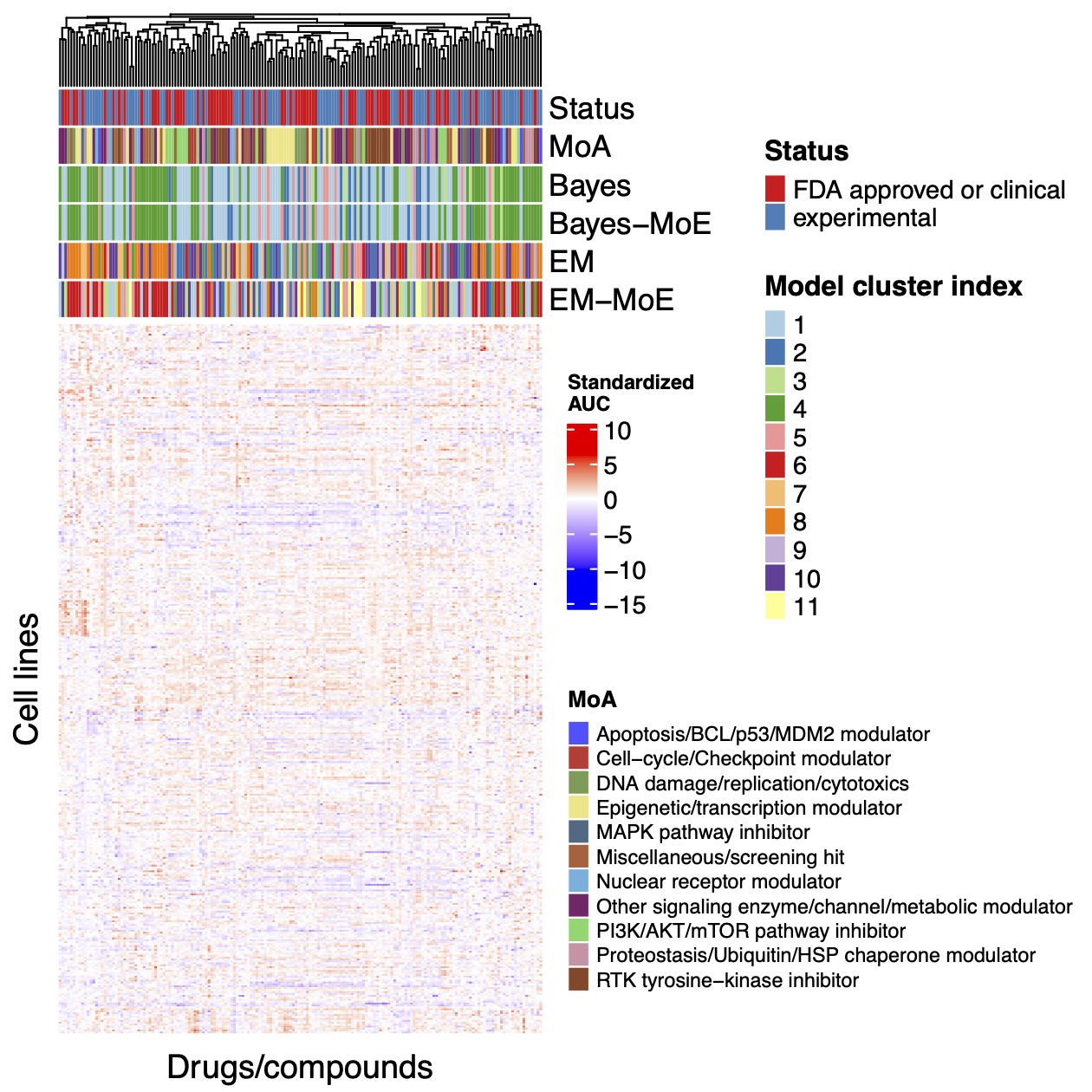}
    \caption{Drug clustering based on sensitivity data. 
    The heatmap shows AUC values for 370 cell lines treated with 172 drugs. AUCs were standardized and subjected to hierarchical clustering (complete linkage method with Euclidean distance) to assess similarity among compounds. The legend ``Model cluster index'' shows the cluster labels from the four methods. As shown, AUC-derived clusters are not aligned with known mechanism-of-actions (\textbf{MoA}), whereas the Bayesian MoE model yields more coherent mechanism-based groupings. 
    }
    \label{fig:realDrug}
\end{figure}

To assess whether the drug groupings produced by the four methods are biologically meaningful, we validated them against well-annotated mechanisms of action (MoA) curated for the same compounds. 
As a baseline, we also compared with hierarchical clustering of drug sensitivity profiles using the area under the dose-response curve (AUC), a common practice in the literature \citep{SeashoreLudlow2015,Zhao2020,Zhao2022,Zeng2022}. 
We restricted the comparison to 172 drugs assayed across 370 cell lines with complete AUC measurements. 
Visual inspection of the dendrogram derived from the standardized AUC values (complete linkage method with Euclidean distance; Figure \ref{fig:realDrug}) indicates that the AUC-based clustering yields groupings that are largely inconsistent with established MoAs and tends to mix agents with disparate targets and pharmacology. 

\begin{figure}[!ht]
    \centering
    \includegraphics[width=13cm]{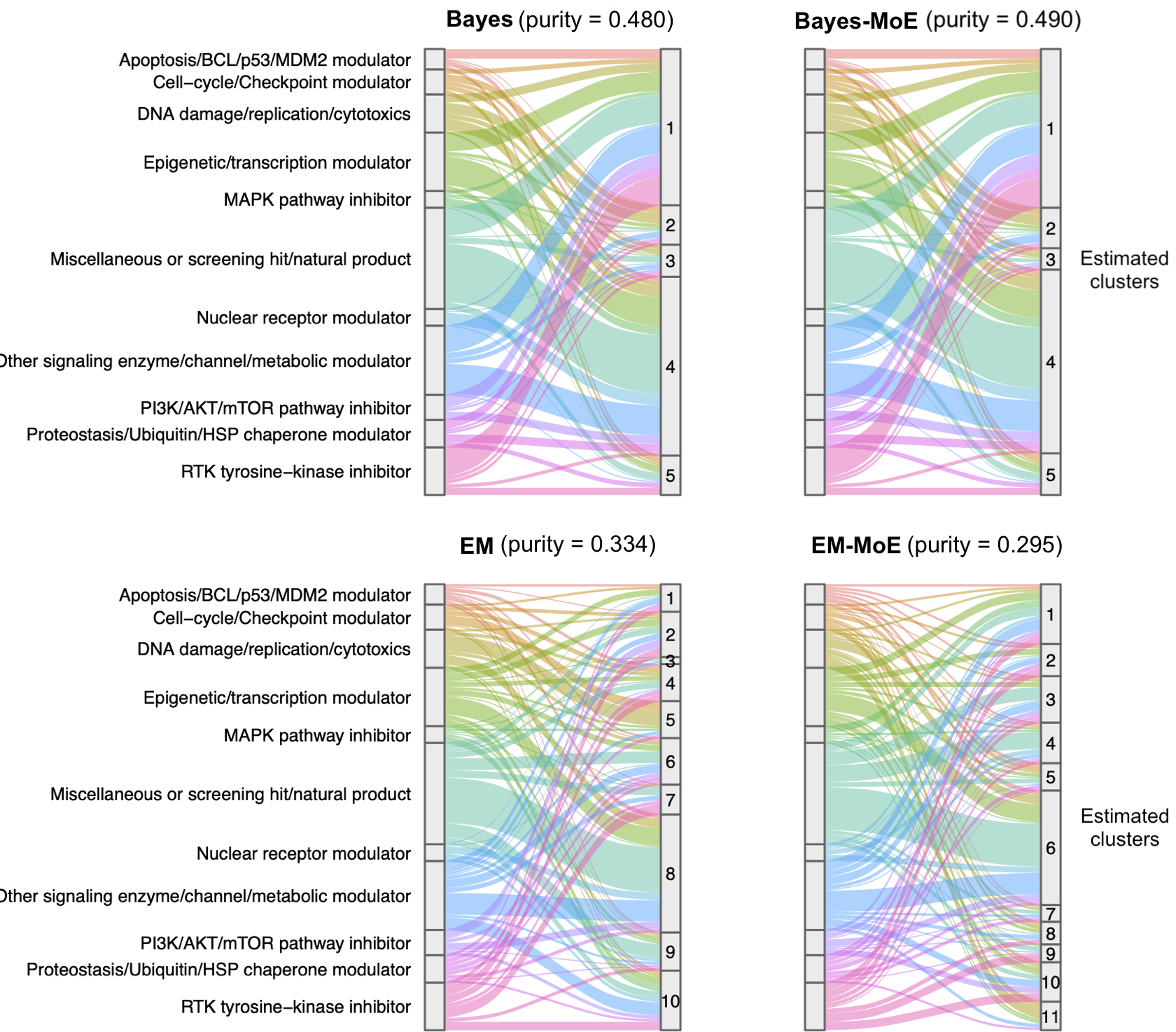}
    \caption{Alluvial diagrams showing flows from MoA groups into individual model-based clusters, illustrating cluster-MoA cohesion. 
    Purity in each panel is defined as: 
    $\text{purity} = \frac{1}{n}\sum_{k=1}^{K} \max_{c} \big| 
    \{ i : \text{cluster}(i)=k \land \text{MoA\_group}(i)=c \} \big|$. 
    }
    \label{fig:realDrug_purity}
\end{figure}

In contrast, the \textbf{Bayes-MoE} model produced clusters that coherently align with biologically interpretable MoAs (purity$=0.490$, Figure \ref{fig:realDrug_purity}). 
We quantify MoA coherence using cluster purity, defined as the fraction of compounds aligned with the majority MoA within each cluster (see the caption of Figure \ref{fig:realDrug_purity} for the exact formula). 
Notably, receptor tyrosine kinase (RTK) inhibitors are consolidated into the same cluster under the \textbf{Bayes-MoE} model for many agents: sunitinib, nilotinib, linifanib, tandutinib, NVP-ADW742, KW-2449 and SU11274. 
This concordance spans multiple RTK classes (e.g., VEGFRs, c-KIT, FLT3), reflecting the shared target class and expected pharmacodynamic similarity. 
For cytotoxic DNA-damaging agents and alkylators, \textbf{Bayes-MoE} similarly aggregates platinum compounds and classical alkylators into coherent clusters consistent with their common MoA: temozolomide, bendamustine, carboplatin and organoplatinum Platin. 
While the Bayesian mixture model (\textbf{Bayes}) often produced similar high-level groupings (purity$=0.480$, Figure \ref{fig:realDrug_purity}), it exhibited inconsistencies for several MoAs. 
For example, histone deacetylase (HDAC) inhibitors were split across multiple clusters under \textbf{Bayes}: vorinostat, belinostat and entinostat — despite sharing a primary target class (HDAC1/2/3/6/8). 
Likewise, classical antimetabolites and nucleoside analogs spanned multiple clusters: methotrexate, gemcitabine, cytarabine, clofarabine and nelarabine, fragmenting a pharmacologically coherent MoA class.

\begin{figure}[!ht]
    \centering
    \includegraphics[width=13cm]{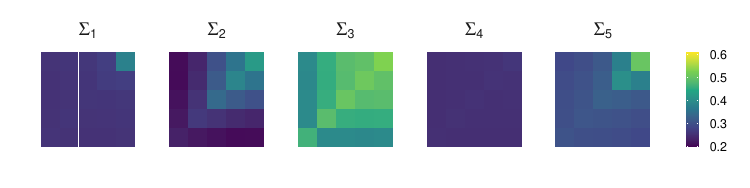}
    \caption{Heatmaps of the posterior mean covariance matrices estimated by \textbf{Bayes-MoE} on the cancer drug screening data with $K=5$. }
    \label{fig_realDrug_Sigmas}
\end{figure}

The EM-based methods (\textbf{EM} and \textbf{EM-MoE}) were even less stable with respect to MoA coherence (purity $0.334$ and $0.295$, respectively; Figure \ref{fig:realDrug_purity}). 
For example, RTK inhibitors were scattered across multiple clusters by \textbf{EM}, and \textbf{EM-MoE} assignments further split the same set of RTK agents across several groups. 
These observations underscore the advantage of \textbf{Bayes-MoE} in capturing modular structure aligned with mechanism-informed biological classes. 
Figure \ref{fig_realDrug_Sigmas} shows the posterior means of the covariance matrices corresponding to the five components.

\section{Discussion and conclusion}
\label{sc_conclusion}

This work develops a unified framework for clustering covariance matrix data.
By formulating a fully Bayesian mixture of Wishart distributions and extending it to a mixture-of-experts architecture, 
we move beyond mean-based clustering paradigms and directly model variability in covariance patterns. 
This perspective is particularly well suited to scientific settings in which dependence structures themselves are the primary objects of interest, such as financial co-movements, functional connectivity in neuroscience, and drug dose–response relationships in pharmacology.

A key contribution of the proposed MoE–Wishart model lies not only in its ability to accommodate covariate-dependent clustering through a probabilistic gating network, but also in the development of tailored inference algorithms that make such modeling practically feasible. By allowing mixing proportions to vary with observed predictors, the model captures systematic changes in latent covariance regimes while maintaining interpretability and flexibility beyond standard mixtures with fixed weights. To support this structure, we develop specialized MCMC algorithms that respect the geometry of symmetric positive definite matrices and yield stable posterior inference, as well as a novel EM procedure for maximum likelihood estimation in the mixture-of-experts setting where closed-form updates are unavailable. Across simulations and real-data applications, this combination of model design and algorithmic innovation leads to improved stability and inferential coherence relative to existing EM-based approaches.

In the cancer drug screening analysis, modeling full dose–dose covariance descriptors preserved rich information on how drug effects co-vary across concentrations and enabled more meaningful drug stratification than scalar summarized descriptor such as AUC. 
Among all methods considered, the Bayesian MoE–Wishart model with 
$K=5$ clusters achieved stable model fit and produced clusters that aligned closely with curated biochemical mechanisms of drug action, 
coherently grouping RTK inhibitors and DNA-damaging or alkylating agents according to their shared pharmacology. 
In contrast, EM-based mixture approaches yielded fragmented and biologically inconsistent groupings. 
These results demonstrate that jointly modeling drug dose–response covariance structure and drug-level covariates within a Bayesian MoE framework provides a robust and interpretable approach for uncovering mechanism-informed drug classes in large-scale screening data.

Several limitations point to promising directions for future research. First, the Wishart likelihood imposes restrictive assumptions on covariance structure, such as unimodality and limited tail behavior, which may be inadequate in settings with extreme dependence patterns or heavy-tailed variability. Extensions to more flexible matrix-variate distributions or nonparametric Bayesian mixtures could alleviate these constraints. Second, scalability remains a challenge for very high-dimensional covariance matrices or large numbers of components, motivating the development of more efficient sampling schemes or variational approximations. Finally, richer gating networks—potentially incorporating nonlinear or hierarchical covariate effects—could further enhance the expressiveness of the MoE–Wishart framework and broaden its applicability to increasingly complex structured data. Theoretical properties would also be the objective of future works \citep{mai2025concentration,mai2025properties,mai2025hightobit,mai2025handling,mai2024high,mai2024concentration}.

Although the MoE framework incorporates covariates at the covariance-descriptor level, it abstracts away individual-level information (e.g., patient-specific clinical or genomic covariates). One natural extension is supervised hierarchical Gaussian process (GP) functional regression, where covariates govern the mean and a GP captures smoothness over the input domain, coupled with residual covariance modeling (e.g., Wishart) and the latent clusters to encode subpopulations similarity. Alternatively, mixtures can be placed directly on GPs, for example, using spectral mixture kernels \citep{Parra2017}, mixtures of GPs \citep{Tresp2000}, or mixtures of GP experts \citep{Rasmussen2001}, to model multimodal, heterogeneous functional responses with complex covariate effects.

\subsubsection*{Acknowledgments}
The views, findings, and opinions presented in this work are exclusively those of the author and do not reflect the official stance of the Norwegian Institute of Public Health. 
Z.Z. was supported by the ERA PerMed under the ERA-NET Cofund scheme of the European Union's Horizon 2020 research and innovation framework program (grant ‘SYMMETRY’ ERAPERMED2021-330).

\subsubsection*{Conflicts of interest/Competing interests}
The authors declare no potential conflict of interests.


\appendix

\renewcommand\thefigure{S\arabic{figure}}    
\renewcommand\thetable{S\arabic{table}}    

\section{Details on Algorithm}\label{secAppendix1}

In Section 3.1 Gibbs-within-MH sampler for mixture model, we can integrate out $\bm{\pi}$ in Eq \eqref{eq_conditional_of_z_i}:
\begin{align*}
&p(z_i=k\mid z_{-i}, S_i,\{\Sigma_\ell,\nu_\ell\},\bm\alpha) \\
\propto &
\int \pi_k f_{\mathcal W}(S_i\mid \nu_k,\Sigma_k) \cdot \text{Dir}(\bm\pi|z_{-i}, \bm\alpha) d\bm\pi \\
\propto &
\int \pi_k f_{\mathcal W}(S_i\mid \nu_k,\Sigma_k) \cdot \text{Dir}(\bm\pi|\bm\alpha) \cdot 
\text{Cat}(z_{-i}|\bm\alpha) d\bm\pi \\
\propto &
\int \pi_k f_{\mathcal W}(S_i\mid \nu_k,\Sigma_k) \cdot \text{Dir}(\bm\pi|\bm a) d\bm\pi, \; \; \text{where }a_j=\alpha_j+n_{-i,j}, \; n_{-i,j}:=\#\{l\neq i | z_l=j\} \\
=&
f_{\mathcal W}(S_i\mid \nu_k,\Sigma_k) 
\int \pi_k\prod_{j=1}^K\frac{\pi_j^{a_j-1}}{B(\bm a)} d\bm\pi\\
=&
f_{\mathcal W}(S_i\mid \nu_k,\Sigma_k) \frac{1}{B(\bm a)}
\int \pi_1^{a_1-1}\pi_2^{a_2-1}\cdots \pi_k^{a_1}\cdots\pi_K^{a_K-1} d\bm\pi \\
=&
f_{\mathcal W}(S_i\mid \nu_k,\Sigma_k) 
\frac{1}{B(\bm a)}B(a_1,...,a_k+1,...,a_K)
\int d\mathbb P_{\bm\pi\sim\text{Dir}(a_1,...,a_k+1,...,a_K)}\\
=&
f_{\mathcal W}(S_i\mid \nu_k,\Sigma_k) \cdot a_k/\sum_{j=1}^Ka_j\\
\propto&
f_{\mathcal W}(S_i\mid \nu_k,\Sigma_k) \cdot (\alpha_k+n_{-i,k})
\end{align*}
This induces better MCMC mixing than sampling from the conditional distribution in Eq \eqref{eq_conditional_of_z_i}.

\section{Additional simulation results}

\begin{table}[!htbp]
\centering
\caption{Effective sample sizes (ESS) for the mixture and mixture-of-experts working models.
Data are generated from the mixture-of-experts model. 
Reported are the means (standard deviations) of ESS over 100 simulations for selected parameters. 
$\nu_1$ and $\nu_2$ are the degrees of freedom of the Wishart distributions for the first two clusters. $\Sigma_{1,11}$ and $\Sigma_{2,11}$ are the $(1,1)$-entries of the corresponding scale matrices. $\beta_{11}$ and $\beta_{12}$ are the drug-status effects for the first two clusters. 
\medskip\label{tab:simMoE_ESS} }
\medskip
  \small
\begin{tabular}{r cc c cc c cc}  
\toprule
\multicolumn{1}{l}{\textbf{Working model}} 
  & \multicolumn{2}{c}{$\bm\nu$} 
  && \multicolumn{2}{c}{$\Sigma$} 
  && \multicolumn{2}{c}{$\beta$} 
  \\
\cmidrule{2-3}
\cmidrule{5-6} 
\cmidrule{8-9} 
 & $\nu_1$ & $\nu_2$ &
 & $\Sigma_{1,11}$ & $\Sigma_{2,11}$ &
 & $\beta_{11}$ & $\beta_{12}$  
 \\
\midrule 
\multicolumn{1}{l}{\textbf{Mixture model}: $p=2$ } 
\medskip\\
$n=200$ &
723 (203) & 169 (53) && 1179 (435) & 216 (78) && - & - 
\\ 
$n=500$ & 
784 (113) & 169 (37) && 1243 (343) & 214 (54) && - & - 
\\
$n=1000$ & 
790 (124) & 169 (28) && 1155 (317) & 200 (40) && - & - 
\bigskip\\
\multicolumn{1}{l}{\textbf{Mixture model}: $p=8$ } 
\medskip\\
$n=200$ & 
1397 (92) & 722 (57) && 8885 (743) & 3222 (426) && - & - 
\\ 
$n=500$ & 
1484 (72) & 814 (54) && 8908 (829) & 3591 (304) && - & - 
\\
$n=1000$ & 
1450 (85) & 761 (49) && 9495 (639) & 3453 (346) && - & - 
\\
\\
\midrule 
\multicolumn{2}{l}{\textbf{MoE model}: $p=2$ } 
\medskip\\
$n=200$ & 
996 (168) & 251 (75) && 1289 (483) & 313 (120) && 133 (71) & 131 (75)
\\
$n=500$ &     
1023 (103) & 223 (43) && 1366 (345) & 263 (57) && 164 (40) & 203 (71)
\\
$n=1000$ &     
1009 (82) & 233 (38) && 1438 (312) & 271 (52) && 179 (53) & 187 (48)
\bigskip\\
\multicolumn{2}{l}{\textbf{MoE model}: $p=8$ } 
\medskip\\
$n=200$ &  
1342 (90) & 752 (53) && 8644 (1009) & 3333 (368) && 407 (106) & 418 (111)
\\
$n=500$ &  
1495 (78) & 814 (47) && 8843 (449) & 3475 (416) && 442 (84) & 446 (91)
\\
$n=1000$ &  
1458 (82) & 796 (44) && 9514 (871) & 3588 (329) && 469 (108) & 439 (83)
\medskip\\
\bottomrule
\end{tabular}
\end{table}

\begin{figure}[!ht]
    \centering
    \includegraphics[width=14cm]{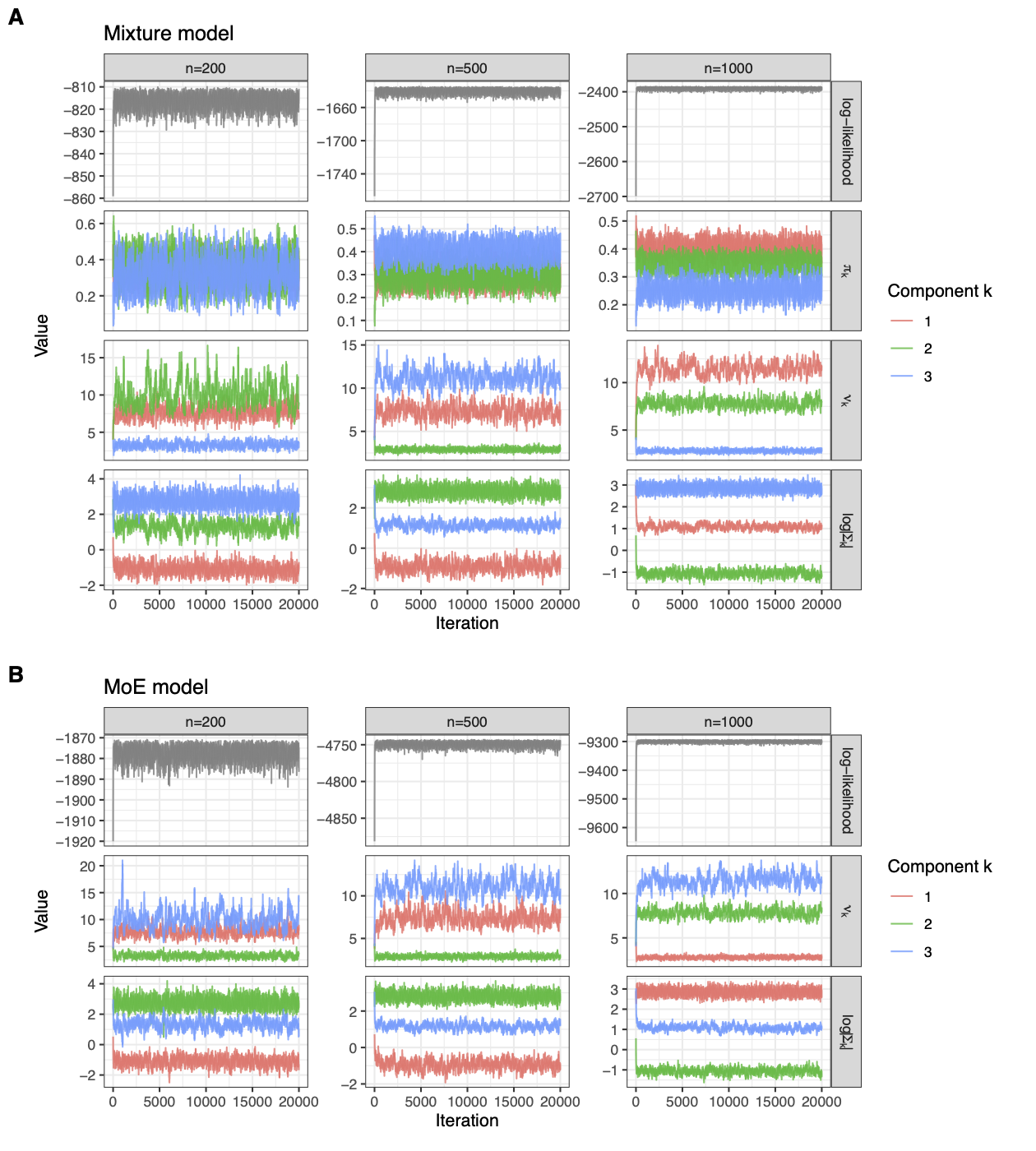}
\caption{Trace plots for MCMC diagnosis for simulated data generated by the \textbf{mixture model} with $\bm{p=2}$ and $n=(200,500,1000)$, fitted by the Bayesian mixture and MoE working models. (A) MCMC diagnosis for the mixture model fit. For each $n$, diagnosis is for the model's log-likelihood, $\pi_k$, $\nu_k$ and $\log|\Sigma_k|$, $k\in\{1,2,3\}$. (B) MCMC diagnosis for the MoE model fit. For each $n$, diagnosis is for the model's log-likelihood, $\nu_k$ and $\log|\Sigma_k|$, $k\in\{1,2,3\}$ . 
}
        \label{figS:simMM}
\end{figure}

\begin{figure}[!ht]
    \centering
    \includegraphics[width=14cm]{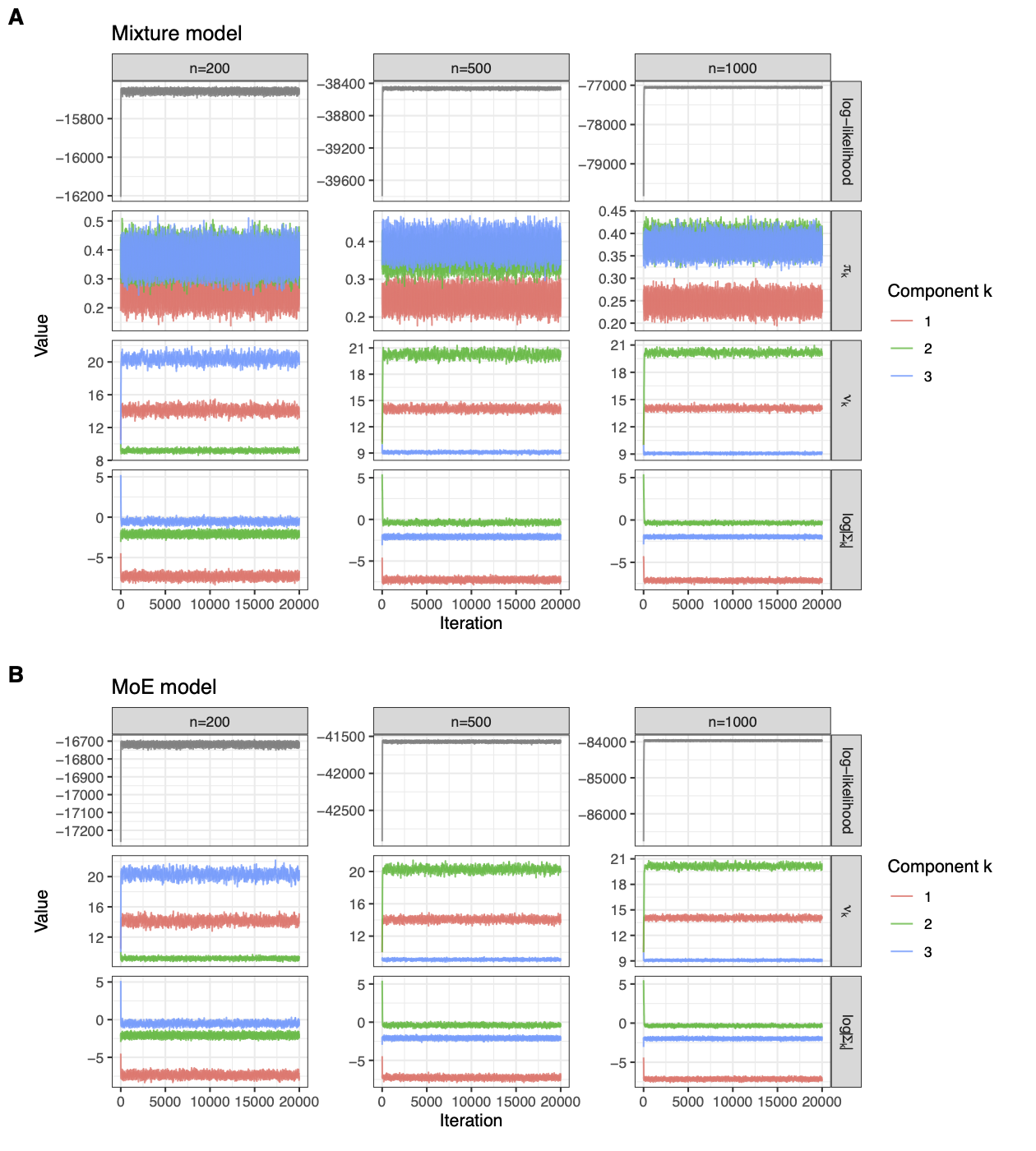}
        \caption{Trace plots for MCMC diagnosis for simulated data generated by the \textbf{mixture model} with $\bm{p=8}$ and $n=(200,500,1000)$, fitted by the Bayesian mixture and MoE working models. (A) MCMC diagnosis for the mixture model fit. For each $n$, diagnosis is for the model's log-likelihood, $\pi_k$, $\nu_k$ and $\log|\Sigma_k|$, $k\in\{1,2,3\}$. (B) MCMC diagnosis for the MoE model fit. For each $n$, diagnosis is for the model's log-likelihood, $\nu_k$ and $\log|\Sigma_k|$, $k\in\{1,2,3\}$.
        }
        \label{figS:simMM_p8}
\end{figure}

\begin{figure}[!ht]
    \centering
   \includegraphics[width=14cm]{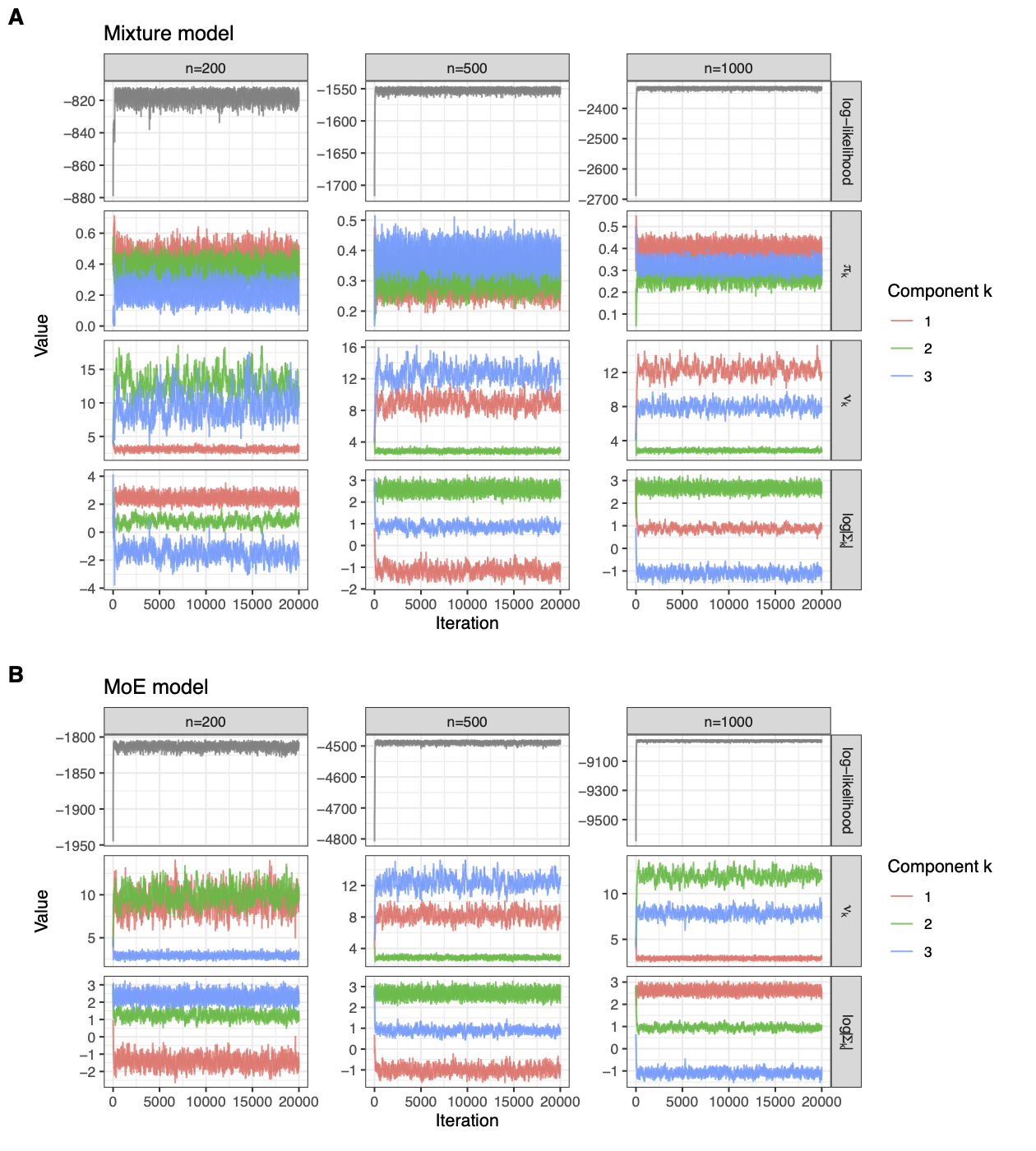}
        \caption{Trace plots for MCMC diagnosis for simulated data generated by the \textbf{MoE model} with $\bm{p=2}$ and $n=(200,500,1000)$, fitted by the Bayesian mixture and MoE working models. (A) MCMC diagnosis for the mixture model fit. For each $n$, diagnosis is for the model's log-likelihood, $\pi_k$, $\nu_k$ and $\log|\Sigma_k|$, $k\in\{1,2,3\}$. (B) MCMC diagnosis for the MoE model fit. For each $n$, diagnosis is for the model's log-likelihood, $\nu_k$ and $\log|\Sigma_k|$, $k\in\{1,2,3\}$.
        }
        \label{figS:simMoE}
\end{figure}

\begin{figure}[!ht]
    \centering
    \includegraphics[width=14cm]{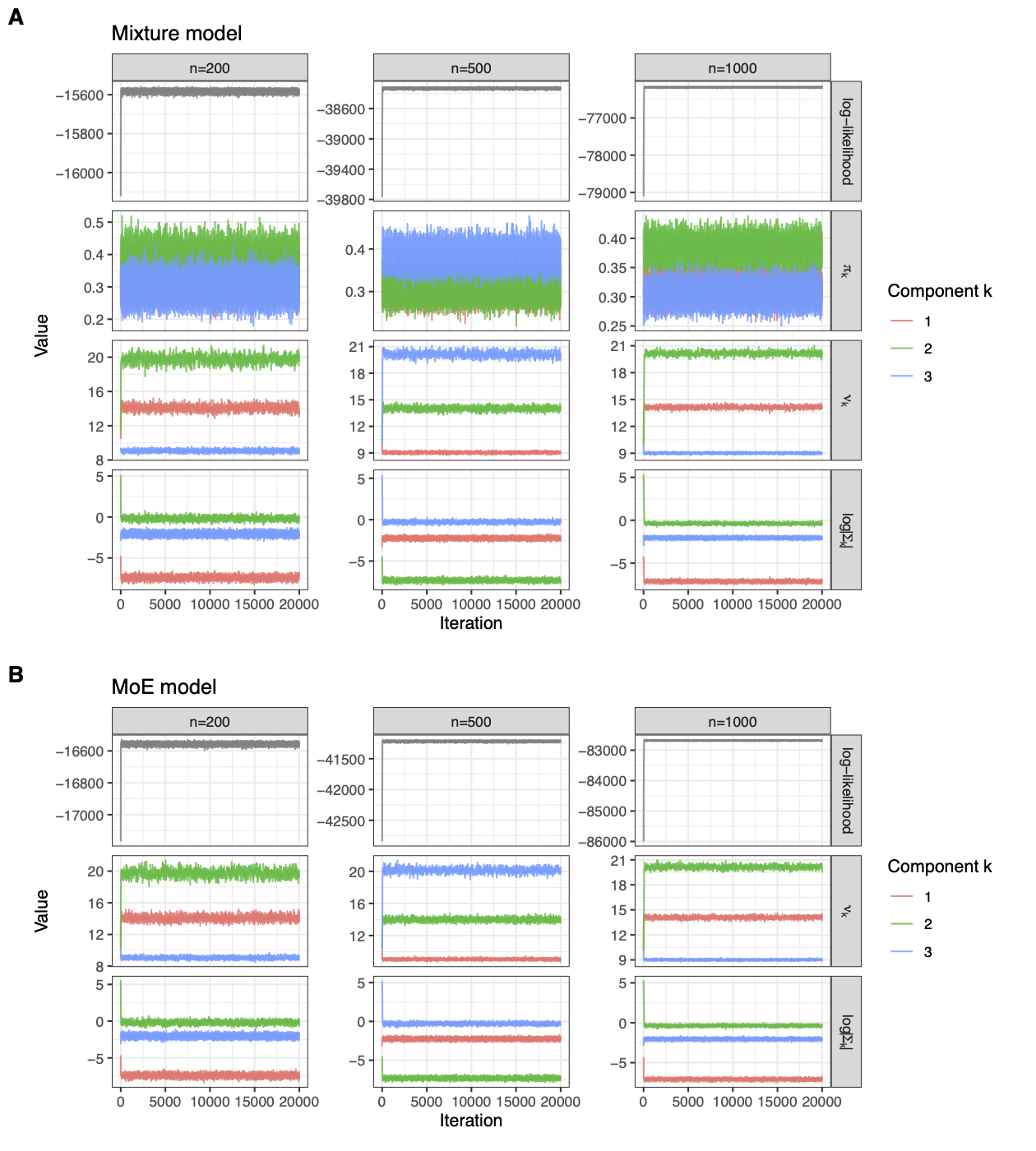}
        \caption{Trace plots for MCMC diagnosis for simulated data generated by the \textbf{MoE model} with $\bm{p=8}$ and $n=(200,500,1000)$, fitted by the Bayesian mixture and MoE working models. (A) MCMC diagnosis for the mixture model fit. For each $n$, diagnosis is for the model's log-likelihood, $\pi_k$, $\nu_k$ and $\log|\Sigma_k|$, $k\in\{1,2,3\}$. (B) MCMC diagnosis for the MoE model fit. For each $n$, diagnosis is for the model's log-likelihood, $\nu_k$ and $\log|\Sigma_k|$, $k\in\{1,2,3\}$.
        }
        \label{figS:simMoE_p8}
\end{figure}

\begin{figure}[htp]
    \centering
    \includegraphics[width=12cm]{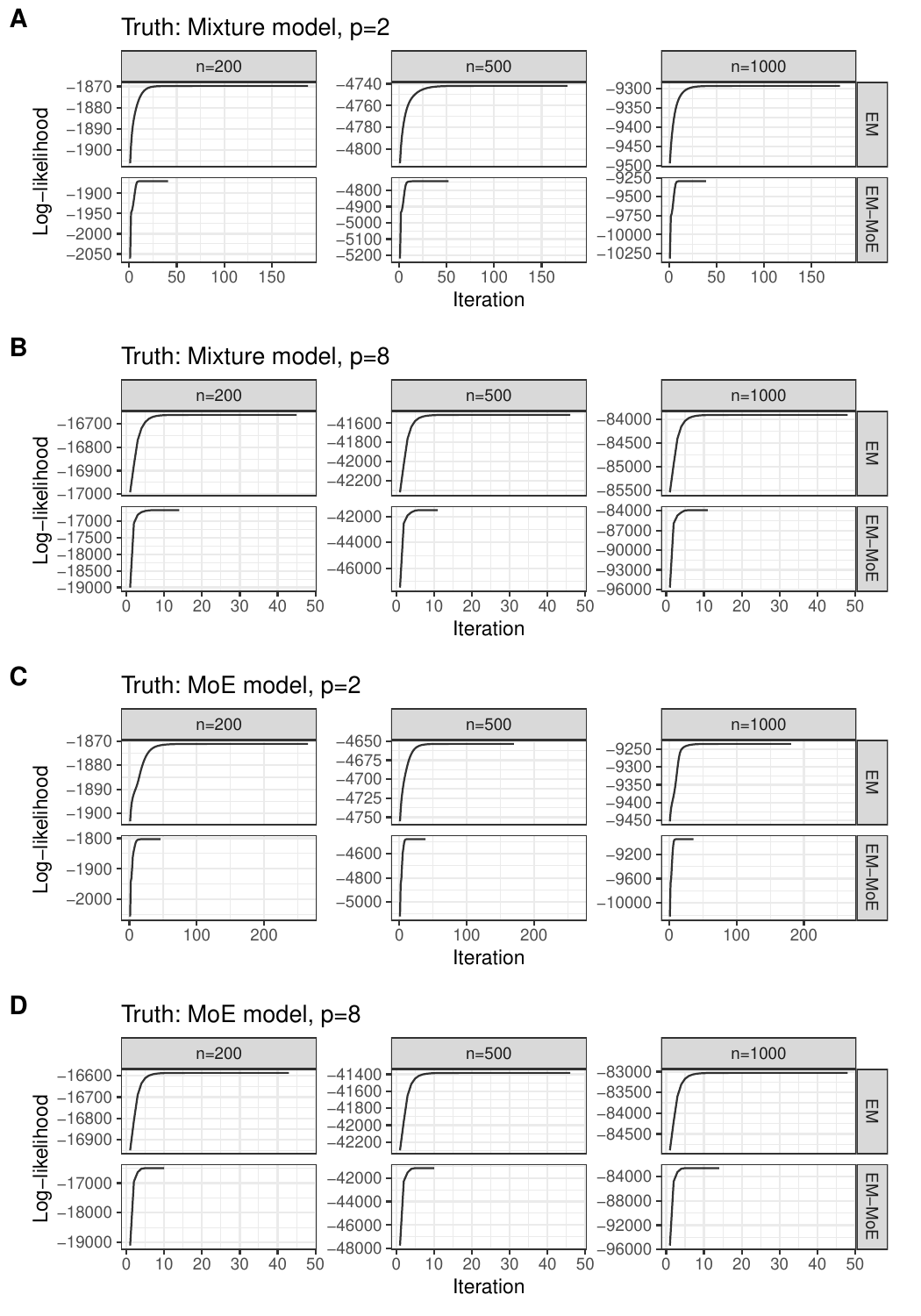}
        \caption{The log-likelihood per iteration by EM algorithms for simulated data generated from the mixture or MoE models (truth). Each of panels (A)-(D) has columns for simulated data $n=(200,500,1000)$, and has rows for mixture and MoE model fits (EM and EM-MoE, respectively). (A) The true underlying model is a mixture model with dimension $p=2$. (B) The true underlying model is a mixture model with dimension $p=8$. (C) The true underlying model is a MoE model with dimension $p=2$. (D) The true underlying model is a MoE model with dimension $p=8$.  }
        \label{figS:simEM}
\end{figure}

\begin{landscape}

\begin{table}[!htp]
\centering
\caption{ 
Simulation results on selection of the number of clusters. Data are generated by the mixture and MoE models with $n=1000$, $p=8$ and true $K=3$. Reported are the mean (standard deviation) of each information criterion over 100 simulations. Bayesian working models are evaluated with elpd$_\text{loo}$ (larger is better) and ICL (smaller is better); EM working models are evaluated with BIC and ICL (both smaller is better). Optimal values per method are bolded. 
\medskip\label{tabS:sim_selectK_n1000_p8} }
\medskip
\begin{tabular}{r ccccc}  
\toprule
\multicolumn{1}{c}{$K$} 
 & 2 & 3 & 4 & 5 & 6 
  \\
\midrule 
\multicolumn{6}{c}{\textbf{Data generated by mixture model}}
  \\
\midrule 
\multicolumn{1}{l}{\textbf{Bayes}} 
\\
elpd$_\text{loo}$ &  
-78662.4 (544.7) & -77428.5 (577.6) & -77428.4 (577.6) & -77428.0 (577.7) & \textbf{-77427.3} (577.6)
\\
ICL & 
157764.8 (1089.5) & \textbf{155549.2} (1154.5) & 155810.7 (1154.5) & 156070.2 (1155.2) & 156329.8 (1155.3)
\smallskip\\
\multicolumn{1}{l}{\textbf{Bayes-MoE}} 
\\
elpd$_\text{loo}$ &  
-85571.1 (544.6) & -84338.2 (577.6) & -84330.3 (577.2) & -84325.7 (577.0) & \textbf{-84320.5} (577.7)
\\
ICL & 
172927.5 (1101.1) & \textbf{171499.7} (1145.8) & 171729.0 (1145.6) & 171961.5 (1144.7) & 172196.3 (1144.5)
\smallskip\\
\multicolumn{1}{l}{\textbf{EM}} 
\\
BIC &  
171506.3 (1089.4) & \textbf{169254.8} (1121.0) & 169338.1 (1157.6) & 169481.5 (1159.5) & 169514.5 (1171.1)
\\ 
ICL &  
171506.4 (1089.5) & \textbf{169276.4} (1120.2) & 169382.0 (1157.4) & 169564.3 (1156.5) & 169586.0 (1168.8)
\\
\multicolumn{1}{l}{\textbf{EM-MoE}} 
\\
BIC &  
171506.4 (1089.4) & \textbf{169230.8} (1155.3) & 169420.4 (1155.2) & 169616.6 (1151.3) & 169802.9 (1155.2)
\\ 
ICL &  
171506.4 (1089.5) & \textbf{169252.6} (1154.6) & 169538.4 (1176.7) & 169794.1 (1164.0) & 170196.2 (1191.2)
\\ 
\\
\multicolumn{6}{c}{\textbf{Data generated by MoE model} }
\\
\midrule 
\multicolumn{1}{l}{\textbf{Bayes}} 
\\
elpd$_\text{loo}$ &  
-77825.9 (508.9) & -76470.2 (537.1) & -76470.1 (537.1) & \textbf{-76469.8} (537.0) & \textbf{-76469.8} (537.0)
\\
ICL & 
156111.9 (1017.9) & \textbf{153674.1} (1073.9) & 153956.7 (1074.0) & 154238.1 (1073.8) & 154521.5 (1074.1)
\smallskip\\
\multicolumn{1}{l}{\textbf{Bayes-MoE}} 
\\
elpd$_\text{loo}$ &  
-84432.7 (506.4) & -82973.6 (538.2) & -82968.6 (538.4) & -82963.9 (537.4) & \textbf{-82958.0} (537.2)
\\
ICL & 
170048.4 (1020.5) & \textbf{168002.7} (1073.5) & 168249.6 (1073.8) & 168503.4 (1072.3) & 168751.4 (1072.0)
\smallskip\\
\multicolumn{1}{l}{\textbf{EM}} 
\\ 
BIC &  
169853.4 (1017.9) & 167874.3 (1490.0) & \textbf{167500.0} (1135.3) & 167628.5 (1104.8) & 167769.9 (1080.3)
\\ 
ICL &  
169853.4 (1017.9) & 167892.0 (1483.6) & \textbf{167537.7} (1133.4) & 167694.8 (1120.2) & 167851.0 (1085.3)
\smallskip\\
\multicolumn{1}{l}{\textbf{EM-MoE}} 
\\ 
BIC &  
169243.1 (1013.0) & \textbf{166530.6} (1076.6) & 166735.7 (1075.0) & 166957.3 (1073.0) & 167152.2 (1073.2)
\\ 
ICL &  
169243.1 (1012.9) & \textbf{166548.0} (1076.7) & 166835.1 (1085.4) & 167112.3 (1089.0) & 167491.9 (1089.9)
\\ 
\bottomrule
\end{tabular}
\end{table}

\end{landscape}
    
\clearpage

\section{Additional results on drug response data analysis}

\begin{table}[!htp]
\centering
\caption{Selection of the number of clusters $K$ on cancer drug response data. Bayesian methods (\textbf{Bayes} and \textbf{Bayes-MoE}) are evaluated with elpd$_\text{loo}$ (larger is better) and ICL (smaller is better). EM methods (\textbf{EM} and \textbf{EM-MoE}) are evaluated with BIC and ICL (both smaller is better). Optimal values per method are bolded. 
\label{tabS:realDrug_selection}    
}
  \footnotesize
\begin{tabular}{r cccccccccc}  
\toprule
\multicolumn{1}{c}{$K$} 
 & 2 & 3 & 4 & 5 & 6 & 7 & 8 & 9 & 10 & 11
  \\
\midrule 
\multicolumn{1}{l}{\textbf{Bayes}} 
\medskip\\
elpd$_\text{loo}$ &  
8269.6 & 9369.8 & 9507.2 & \textbf{9700.2}  & 9696.9 & \textbf{9700.4} & 9698.7  & - & - & - 
\\
ICL & 
-16371 & -18461.6 & -18628.7 & \textbf{-18892.5} & -18775.6 & -18670.7 & -18554.1
\bigskip\\
\multicolumn{1}{l}{\textbf{Bayes-MoE}} 
\medskip\\
elpd$_\text{loo}$ &  
6054.9 &  7158.1 &  7283.5 &  7463.4 &  \textbf{7466.0} &  7462.2 &  7465.1  & - & - & - 
\\
ICL & 
-11560.8 & -13324.8 & -13397.9 & \textbf{-13549.6} & -13438.6 & -13316.6 & -13202.9 & - & - & - 
\bigskip\\
\multicolumn{1}{l}{\textbf{EM}} 
\medskip\\
BIC &  
-12156.4 & -15347.2 & -16034.7 & -16591.1 & -16769 & -17365.2 & -17508.3 & -17359.9 & \textbf{-17742.7} & -17595.2
\\ 
ICL &  
-12139.3 & -15335.2 & -16007.3 & -16555.2 & -16728.2 & -17336.9 & -17473.2 & -17328.8 & \textbf{-17704.9} & -17555.8
\bigskip\\
\multicolumn{1}{l}{\textbf{EM-MoE}} 
\medskip\\
BIC &  
-12167.3 & -15369.4 & -15927.1 & -16646.1 & -17102.4 & -17374.4 & -17516.8 & -17768.2 & -17810.3 & \textbf{-17874.0}
\\ 
ICL &  
-12150.3 & -15357.5 & -15900.4 & -16614.8 & -17072.1 & -17343 & -17479.6 & -17731.4 & -17770.0 & \textbf{-17839.7}
\\ 
\bottomrule
\end{tabular}
\end{table}

\begin{table}[!htbp]
\centering
\caption{ Additional results on cancer drug sensitivity real data.
Parameter estimation by four competing methods. The two mixture models are \textbf{Bayes} (with $K=5$ clusters) and \textbf{EM} (with $K=10$ clusters). The two MoE models are \textbf{Bayes-MoE} (with $K=5$ clusters) and \textbf{EM-MoE} (with $K=11$ clusters), and included two concomitant covariates: (i) drug status coded as a binary variable and (ii) the first principal component from PCA of the drug SMILES-derived fingerprint matrix. 
Selected parameters are reported with point estimates by \textbf{EM} and \textbf{EM-MoE} methods, and with posterior means (95\% credible interval) by \textbf{Bayes} and \textbf{Bayes-MoE}. $\nu_1$ and $\nu_2$ are the first two clusters' Wishart distributions' degrees of freedom. $\Sigma_{1,11}$ and $\Sigma_{2,11}$ are the first two clusters' Wishart distributions' scale matrices' first entries. $\pi_1$ and $\pi_2$ are the first two clusters' mixture weights. $\beta_{11}$ and $\beta_{12}$ are the first two clusters' drug status' effects. 
\label{tabS:realDrug} }
  \small
\begin{tabular}{r cc cr ccc}  
\toprule
\textbf{Parameter} & \textbf{Bayes} & \textbf{EM} 
&
& & \textbf{Bayes-MoE} & \textbf{EM-MoE} 
\\ \midrule
$\nu_1$ & 10.9 (10.3, 11.4) & 31.4 
& &  & 10.8 (10.3, 11.4) & 33.6 
\\
$\nu_2$ & 12.4 (11.0, 13.9) & 37.2 
& &  &  12.4 (10.9, 14.0) & 24.5 
\\
$\Sigma_{1,11}$ & 0.3 (0.24, 0.28) & 0.09 
& & 
& 0.3 (0.24, 0.28) & 0.08 
\\
$\Sigma_{2,11}$ & 0.2 (0.19, 0.28) & 0.08 
& & 
& 0.2 (0.19, 0.28) & 0.11 
\\
$\pi_1$ & 0.4 (0.32, 0.42) & 0.06 
& & $\beta_{11}$ & -0.2 (-0.87, 0.51) & 0.38 
\\
$\pi_2$ & 0.1 (0.05, 0.11) & 0.10
& &  $\beta_{12}$ & -1.0 (-2.01, 0.05) & -0.34
\\
\bottomrule
\end{tabular}
\end{table}

\begin{figure}[!ht]
    \centering
    \includegraphics[width=15cm]{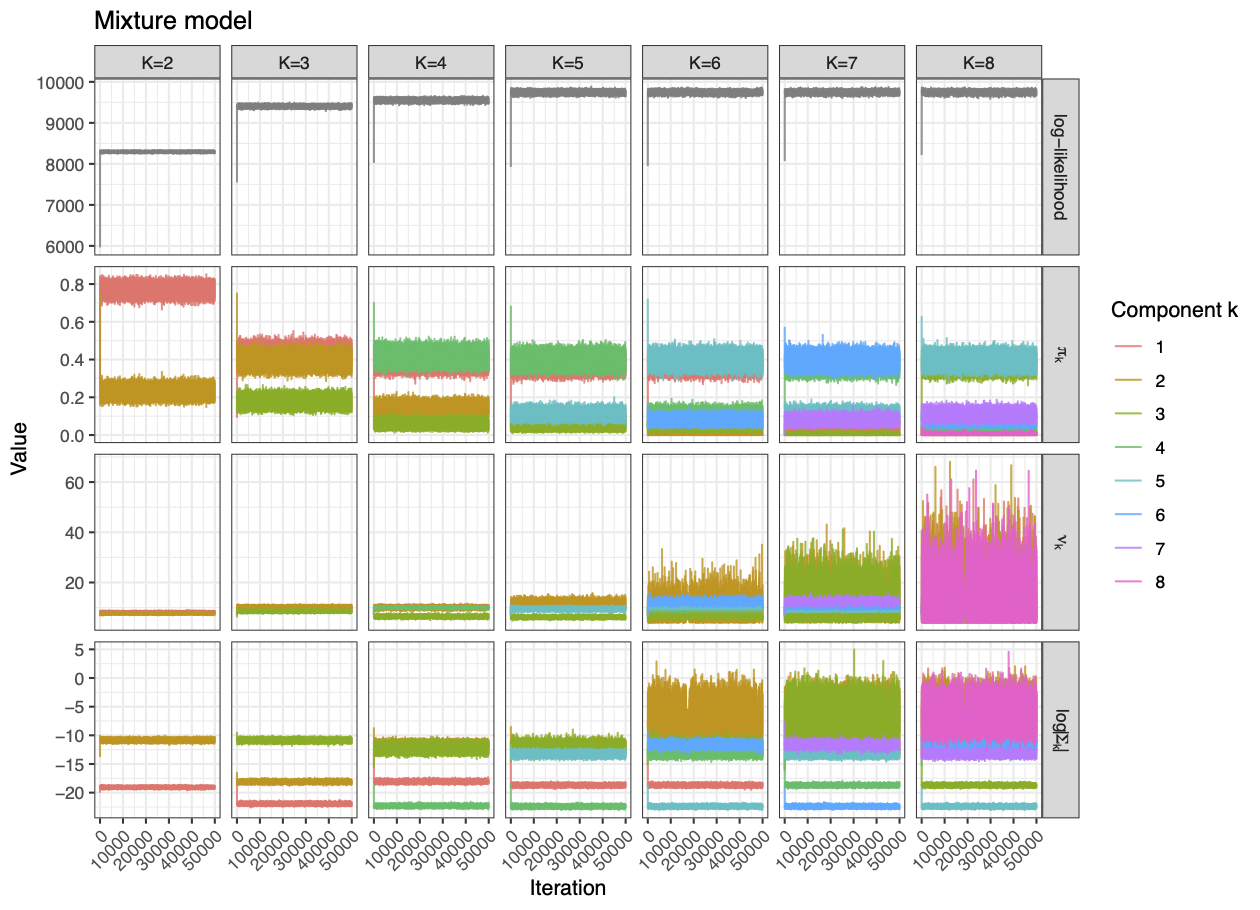}
    \caption{Additional results on cancer drug sensitivity real data.
    Trace plots for MCMC diagnostics of parameters for drug sensitivity data fitted by the \textbf{mixture model} with specified $K\in\{2,3,4,5,6,7,8\}$. 
    The first row shows the trace plots of the model's log-likelihood with respect to $K$. 
    The second row shows the trace plots of $\pi_{k}$ (mixture weight), $k=1,2,...,K$, with respect to $K$. 
    The third row shows the trace plots of $\nu_{k}$ (Wishart distribution's degree of freedom) with respect to $K$. 
    The fourth row shows the trace plots of $\log|\Sigma_k|$ ($\Sigma_k$ is the Wishart distribution's scale matrix) with respect to $K$. 
    The legend shows different colors for component/cluster index $k$.
    }
    \label{figS:realMM}
\end{figure}

\begin{figure}[!ht]
    \centering
    \includegraphics[width=15cm]{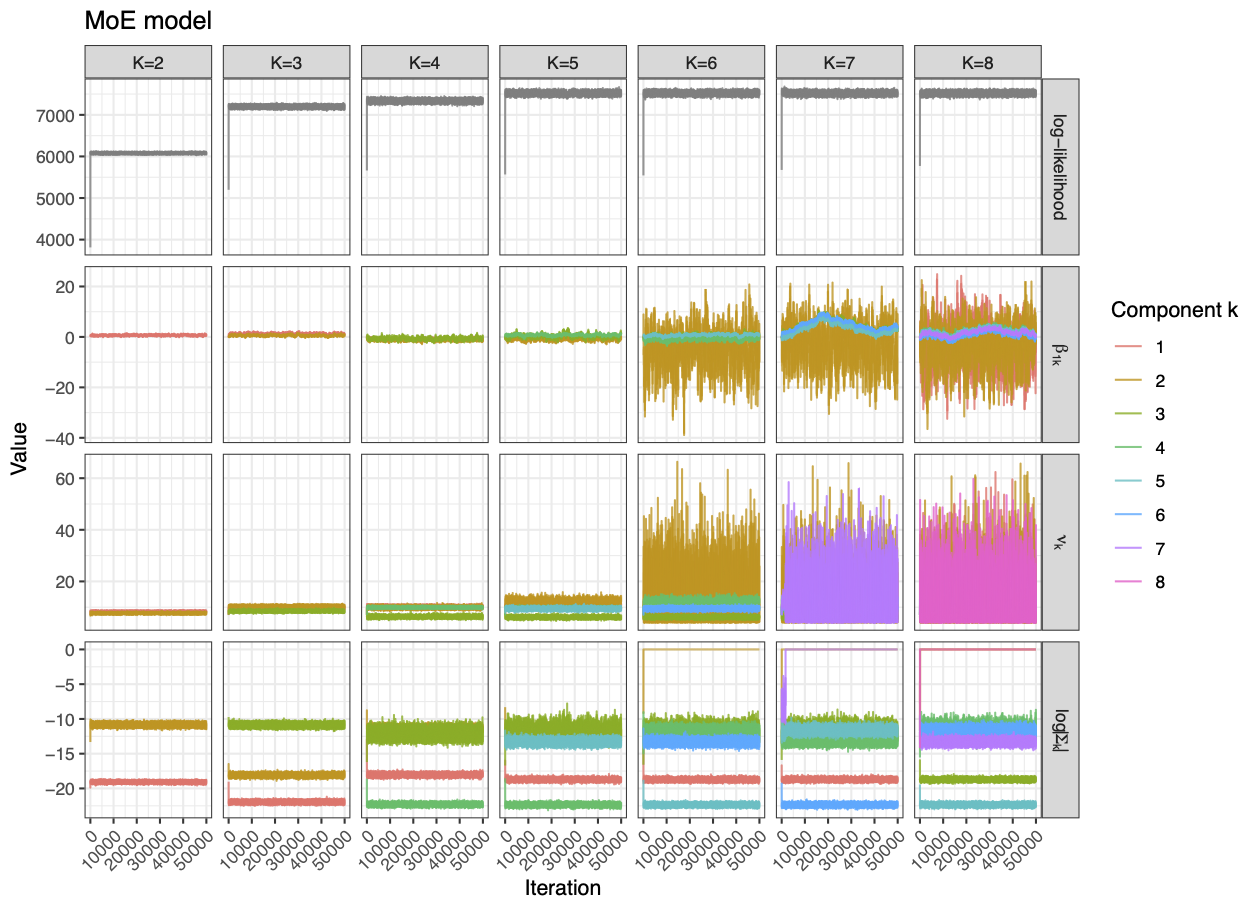}
    \caption{
    Additional results on cancer drug sensitivity real data.
    Trace plots for MCMC diagnostics of parameters for drug sensitivity data fitted by the \textbf{mixture-of-experts model} with specified $K\in\{2,3,4,5,6,7,8\}$. 
    The first row shows the trace plots of the model's log-likelihood with respect to $K$. 
    The second row shows the trace plots of $\beta_{1k}$ (drug-status effect), $k=1,2,...,K$, with respect to $K$. 
    The third row shows the trace plots of $\nu_{k}$ (Wishart distribution's degree of freedom) with respect to $K$. 
    The fourth row shows the trace plots of $\log|\Sigma_k|$ ($\Sigma_k$ is the Wishart distribution's scale matrix) with respect to $K$. 
    The legend shows different colors for component/cluster index $k$.  }
    \label{figS:realMoE}
\end{figure}

\end{document}